\documentclass[11pt]{article}

\usepackage{jheppub}
\usepackage{graphicx}
\usepackage[export]{adjustbox}
\usepackage{subcaption}
\usepackage{color}
\usepackage{tikz}
\usetikzlibrary{positioning, quotes, decorations.markings, decorations.pathmorphing, shapes, calc}
\usepackage{tikz-cd}
\usepackage{slashed}
\usepackage{braket}
\usepackage{comment}
\newcommand{\ol}{\overline}
\newcommand{\wt}{\widetilde}
\newcommand{\Tr}{\mathrm{Tr}}

\newcommand*\diff{\mathop{}\!\mathrm{d}}
\newcommand{\pd}{{\partial}}

\newcommand{\BA}{{\mathbf A}}
\newcommand{\BB}{{\mathbf B}}

\newcommand{\BW}{{\mathbf W}}
\newcommand{\BX}{{\mathbf X}}
\newcommand{\BY}{{\mathbf Y}}

\newcommand{\BPsi}{{\mathbf \Psi}}
\newcommand{\BPhi}{{\mathbf \Phi}}
\newcommand{\BPi}{{\mathbf \Pi}}

\newcommand{\BLambda}{{\mathbf \Lambda}}
\newcommand{\Bmu}{{\mathbf \mu}}
\newcommand{\BOmega}{{\mathbf \Omega}}

\newcommand{\C}{\mathbb{C}}
\newcommand{\R}{\mathbb{R}}
\newcommand{\Z}{\mathbb{Z}}

\newcommand{\PP}{\mathbb{P}}
\newcommand{\PT}{\mathbb{PT}}

\newcommand{\fsl}{\mathfrak{sl}}

\newcommand{\fg}{\mathfrak{g}}

\newcommand{\m}{\mathfrak{m}}

\newcommand{\CA}{{\mathcal A}}
\newcommand{\CB}{{\mathcal B}}

\newcommand{\CD}{{\mathcal D}}
\newcommand{\CE}{{\mathcal E}}

\newcommand{\CG}{{\mathcal G}}

\newcommand{\CK}{{\mathcal K}}
\newcommand{\CL}{{\mathcal L}}
\newcommand{\CM}{{\mathcal M}}
\newcommand{\CN}{{\mathcal N}}
\newcommand{\CO}{{\mathcal O}}

\newcommand{\CV}{{\mathcal V}}
\newcommand{\CW}{{\mathcal W}}  
\newcommand{\CX}{{\mathcal X}}
\newcommand{\CY}{{\mathcal Y}}
\newcommand{\CZ}{{\mathcal Z}}

\newcommand{\norm}[1]{{{:\!{#1}\!:}}}
\newcommand{\be}{\begin{equation}}
	\newcommand{\ee}{\end{equation}}

\title{Twistorial monopoles \& chiral algebras}
\author[1]{Niklas Garner}
\emailAdd{nkgarner@uw.edu}
\author[1]{Natalie M. Paquette}
\emailAdd{npaquett@uw.edu}
\affiliation[1]{Department of Physics, University of Washington, Seattle, USA}

\abstract{We initiate the study of how the insertion of magnetically charged states in 4d self-dual gauge theories impacts the 2d chiral algebras supported on the celestial sphere at asymptotic null infinity, from the point of view of the 4d/2d twistorial correspondence introduced by Costello and the second author. By reducing the 6d twistorial theory to a 3d holomorphic-topological theory with suitable boundary conditions, we can motivate certain non-perturbative enhancements of the celestial chiral algebra corresponding to extensions by modules arising from 3d boundary monopole operators. We also identify the insertion of 4d (non-abelian) monopoles with families of spectral flow automorphisms of the celestial chiral algebra.}

\begin{document}
\maketitle


\section{Introduction}\label{sec:intro}

It has been recently appreciated that four-dimensional quantum field theories, possibly coupled to gravity, on asymptotically flat spacetimes have classical asymptotic symmetries which organize into chiral algebras \cite{Guevara:2021abz, Strominger:2021lvk}. The operator product on algebra elements arises from certain collinear limits of scattering amplitudes of massless states. Utilizing the standard map between 4d null momenta and points on the sphere at asymptotic null infinity invites a possible interpretation of the chiral algebra as the holomorphic symmetry algebra of an exotic two-dimensional conformal field theory supported on the celestial sphere, which is proposed to be holographically dual to the four-dimensional ``bulk'' theory. 

The impact of quantum corrections on these classical chiral algebras, and their four-dimensional origin, has been recently explored by several groups, e.g. \cite{Ball:2021tmb, Bhardwaj:2022anh, Monteiro:2022lwm, CP22, CPassoc, Fernandez:2023abp, Bittleston:2022jeq}. It was observed in \cite{CPassoc} that in a formulation of self-dual Yang Mills (SDYM) theory, the chiral algebra of asymptotic symmetries, when deformed to include the theory's one-loop collinear splitting amplitude, failed to be associative. This ``quantum'' failure of associativity was directly related to an obstruction to lifting the four-dimensional theory to a local holomorphic theory on twistor space. Such an obstruction manifests as a gauge anomaly in the six-dimensional holomorphic theory \cite{Costello:2021bah}. Conversely, one can couple SDYM to a certain scalar field with a quartic kinetic term, which cancels the 6d gauge anomaly via a Green-Schwarz-like mechanism on twistor space and obtain an associative chiral algebra, including loop effects \cite{CP22, CPassoc, Fernandez:2023abp}. In this work we focus on theories with associative chiral algebras: either self-dual theories at tree-level or, when we wish to move beyond classical physics, such integrable ``twistorial'' 4d theories, with lifts to local, non-anomalous holomorphic theories on twistor space, guaranteeing chiral algebra associativity. Such theories can sometimes be uplifted to genuine topological string theories \cite{Costello:2019jsy}, and produce integrable toy models of flat space holography \cite{Costello:2022jpg}, along the lines of the twisted holography program \cite{Costello:2018zrm, CPKoszul}. It will be interesting to understand if associativity can be restored by other means, or if such chiral algebra axioms can be somehow relaxed to accommodate an exotic CFT dual, perhaps along the lines of \cite{Bakalov_2003}. 

Among the many open questions relating to the physics of celestial CFTs and their associated chiral algebras is how to formulate the 2d avatar of magnetically charged states in four dimensions. Soft theorems in (abelian) gauge theories with magnetically charged objects have been studied in \cite{Strominger:2015bla, Kapec:2021eug}. In this work, we will instead emphasize the chiral algebraic formulation of conformally soft modes of such states, leveraging the twistorial construction of \cite{CP22}. Along the way, we will unearth a host of novel or under-explored phenomena in the representation theory of non-unitary 2d chiral algebras.

To proceed, let us recall part of the basic setup of \cite{CP22}. Twistor space is a complex manifold given by the nontrivial fibration
\begin{equation}
	\mathbb{PT}:= \mathcal{O}(1)\oplus \mathcal{O}(1) \rightarrow \mathbb{CP}^1
\end{equation}
i.e. $\PT$ is the total space of two copies of the line bundle $\CO(1) \to \PP^1$, and as a \textit{real} manifold is equivalent to
\begin{equation}
	\mathbb{PT} \simeq \mathbb{R}^4 \times \mathbb{CP}^1.
\end{equation}
We may study various local, anomaly-free holomorphic theories on $\mathbb{PT}$, which reduce to the 4d theories of interest upon dimensionally reducing along the $\mathbb{CP}^1$. In particular, we focus on the 6d theories studied in \cite{Costello:2021bah} given by holomorphic BF theories with particular gauge algebras $\mathfrak{g}$, each coupled to a free limit of the Kodaira-Spencer theory of ``gravity'' (complex structure deformations) \cite{Bershadsky:1993cx} via a holomorphic Chern-Simons term with a carefully chosen coefficient. These reduce to non-unitary CFTs in 4d, given by SDYM theories coupled to a quartic axion-like field. The 2d chiral algebra of asymptotic symmetries was constructed by twistorial methods explained in \cite{CP22}, and is supported on the zero section of the twistor fibration, which in turn can be identified with the celestial sphere at null infinity. 

An alternative view of this chiral algebra presented in \cite{CP22}, which is well-suited to our present purposes, is as follows. Consider $\mathbb{PT} \backslash \mathbb{CP}^1 \simeq S^3 \times \mathbb{R}_{>0} \times \mathbb{CP}^1$, where we remove the sphere corresponding to the origin of $\mathbb{R}^4$. We consider an alternative compactification of the 6d theory, down to 3d dimensions, by reducing along $S^3$; in contrast to the usual reduction along $\mathbb{CP}^1$, there is an infinite tower of KK modes that must be included to access all the (infinitely many) states in the vacuum module of the chiral algebra. Compactifying a 6d holomorphic ``gravitational'' theory in this manner yields a 3d holomorphic-topological \textit{non}-gravitational theory on $\mathbb{R}_{>0} \times \mathbb{CP}^1$, where the chiral algebra furnishes the algebra of boundary local operators supported on the holomorphic $\mathbb{CP}^1$ at infinity in the radial coordinate. 3d holomorphic-topological theories are known to arise from certain supersymmetric twists of 3d $\mathcal{N}=2$ theories \cite{ACMV} and in fact, it is easy to identify ``downstairs'' which 3d $\mathcal{N}=2$ theory can be twisted to yield the resulting holomorphic-topological theory with the ``celestial'' chiral algebra as its boundary chiral algebra \cite{CP22}. This 3d perspective will be our primary window into the physics of magnetic states and their impact on the chiral algebra. 

On the other hand, three-dimensional $\mathcal{N}=2$ gauge theories have long been intense objects of study. They often flow to strongly interacting superconformal field theories and enjoy a host of infrared dualities, as well as qualitative similarities to their non-supersymmetric cousins with useful applications to condensed matter physics. Monopoles in these 3d theories have been studied from many points of view (see, e.g., the classic papers \cite{Aharony:1997bx, Intriligator:2013lca}). Particularly important for our purposes, 1/2-BPS boundary conditions of these theories have been studied extensively in \cite{Gadde:2013wq, Gadde:2013sca, Okazaki:2013kaa, DGP}, and in \cite{DGP} it was shown that Dirichlet boundary conditions on the gauge field support nontrivial boundary monopole operators as well. Such 1/2-BPS boundary conditions are compatible with the holomorphic-topological twist, and may be studied directly in the first-order formalism from this point-of-view \cite{CDG}. Further, the resulting bulk (associative, commutative) chiral algebras of local operators, including monopoles, and boundary (associative, non-commutative) chiral algebras, which may also include monopoles, were explored in \cite{CDG}. In non-abelian gauge theories, boundary monopoles remain difficult to work with directly\footnote{Geometrically, one must exhibit a chiral algebra structure on the Dolbeault homology of the affine Grassmannian $\textrm{Gr}_G$ associated to the gauge group $G$, with values in a certain line bundle; for non-abelian $G$, the non-trivial geometry of $\textrm{Gr}_G$ makes it difficult to make the chiral algbra structure explicit.} but 3d abelian bulk and boundary monopoles can be accessed concretely in the twisted formalism \cite{Zeng:2021zef}. The primary purview of this note is to study monopoles in 3d holomorphic-topological theories which support ``celestial'' chiral algebra boundary conditions. By pulling these states back to twistor space and then ``pushing'' them back down to four-dimensions, we can explore various aspects of the flat space holographic dictionary including the corresponding nonperturbative field configurations.

The plan for the rest of this note is as follows. In Section \ref{sec:monopoles}, we review the physics of 3d monopoles in holomorphic-topologically twisted theories, and how monopoles localized on holomorphic boundaries implement spectral flow automorphisms in perturbative boundary chiral algebras, at least in the abelian setting. We also emphasize how boundary monopoles enforce integrality of the charge of modules of the perturbative chiral algebra (in appropriate conventions), and impose relations among such modules. Classes of such modules are realized physically in 3d theories with boundary via Wilson lines ending on the boundary chiral algebra plane, and we construct, via Koszul duality, the analogous (holomorphic) Wilson line modules on twistor space associated to the perturbative celestial chiral algebra in \ref{sec:KDmodules}. These modules will interact with, and be constrained by, the celestial boundary monopoles that generate the spectral flow automorphisms.

In Section \ref{sec:celestial} we review the uplift of 4d self-dual gauge theories to 6d local holomorphic theories, and in turn recall how the latter can also be viewed as 3d holomorphic-topological theories with boundary, with an infinite number of fields, following \cite{CP22}. We discuss possible nonperturbative extensions of perturbative 3d boundary algebras (in this context, equivalently, 4d celestial chiral algebras) in self-dual QED, and we further discuss 4d abelian self-dual magnetically charged states and some puzzles that arise when trying to map such states to 3d boundary algebras. Finally, in Section \ref{sec:nonabelian} we study self-dual non-abelian gauge theories, and show how 4d magnetically charged states lead to spectral flow automorphisms on the celestial chiral algebra. The spectral flow automorphisms on the celestial chiral algebra (and their inverses) arising from the insertion of a 4d self-dual magnetically charged state are presented in \eqref{eq:nullSF}, \eqref{eq:invnullSF}. We then conclude with open questions and future directions.

To aid in readability, we include the following table of relevant mathematical symbols. In this table $\CM$ is a 3-manifold with a THF structure; $G$ is an algebraic group (over $\C$); and $X$ is a complex manifold with Dolbeault operator $\overline{\pd}$ and $\CE$ is a holomorphic vector bundle over $X$.

\begin{table}[h!]
	\centering
	\begin{tabular}{c|c}
		$\BOmega = \Omega(\CM)/(\diff z)$ & DG algebra of forms on $\CM$ modulo forms proportional to $\diff z$\\
		$\BOmega^{(j)}$ & $\BOmega$ twisted by $\diff z^j$\\
		$\Pi V$ & parity shift operator of a super vector space $V$\\
		$\CO = \C[\![z]\!]$ & algebra of formal series in a variable $z$\\
		$\mathbb{D} = \textrm{Spec}\CO$ & the formal disk\\
		$G_c$ & maximal compact subgroup of $G$\\
		$G_\CO = \textrm{Maps}(\mathbb{D}, G)$ & $\CO$-valued elements of $G$\\
		$\CK = \C(\!(z)\!)$ & algebra of formal Laurent series in a variable $z$\\
		$\overset{\circ}{\mathbb{D}} = \textrm{Spec}\CK$ & the formal punctured disk\\
		$G_\CK = \textrm{Maps}(\overset{\circ}{\mathbb{D}}, G)$ & $\CK$-valued elements of $G$\\
		$\mathbb{B} = \mathbb{D} \cup_{\overset{\circ}{\mathbb{D}}} \mathbb{D}$ & the formal bubble\\
		$\PP^n$ & $n$-dimensional projective space\\
		$\CO(m) \to \PP^n$ & invertible line bundles on $\PP^n$\\
		$\PT = \CO(1)\oplus \CO(1) \to \PP^1$ & 4 dimensional twistor space\\
		$H^\bullet_{\overline{\pd}}(X,\CE) = \bigoplus_i H^i_{\overline{\pd}}(X, \CE)$ & Dolbeault cohomology of $X$ with values in $\CE$\\
		$H_{\bullet, \overline{\pd}}(X,\CE) = \bigoplus_i H_{i,\overline{\pd}}(X, \CE)$ & Dolbeault homology of $X$ with values in $\CE$\\
	\end{tabular}
	\caption{Collection of commonly used mathematical symbols and their meaning.}
	\label{table:symbols}
\end{table}

\section{Monopoles in 3d Gauge Theories}
\label{sec:monopoles}
We begin by reviewing basic aspects of monopoles in 3d $\mathcal{N}=2$ gauge theories, focusing mostly on abelian gauge theories, as well as aspects of boundary monopoles on Dirichlet boundary conditions. For more details on the latter, see \cite{DGP, CDG}.

Local operators in 3d gauge theories have a natural flavor symmetry, called the topological flavor symmetry, with an associated conserved charge called monopole number. This monopole number is determined by surrounding the given local operator by a sphere $S^2$ and measuring the topological type of the gauge field sourced by this operator. If $G_c$ is the (compact, connected) gauge group, a topological classification of principal bundles on $S^2$ is given by $\pi_1(G_c)$: if we cover $S^2$ with two charts covering the northern and southern hemispheres, the transition function on the overlap is topologically equivalent to a map from the equatorial $S^1$ to $G_c$. Local operators with non-trivial monopole number are known as monopole operators.%
\footnote{We note that there are local operators with trivial monopole number in non-abelian gauge theories that still warrant the name monopole operator, e.g. monopoles in gauge theories with simply connected gauge group.} %

\subsection{Monopoles in $HT$-twisted $\CN=2$}
\label{sec:N=2HT}
A particularly important class of monopoles for our later applications are those that arise in twisted $\CN=2$ gauge theories. Recall that the 3d $\CN=2$ supersymmetry algebra has two fermionic generators $Q_\alpha, \ol{Q}_\alpha$ that transform as spinors with respect to $\textrm{Spin}(3) \cong SU(2)$; their anti-commutators (in the absence of central charges) take the form
\be
	\{Q_\alpha, \ol{Q}_\beta\} = (\sigma^\mu)_{\alpha \beta} P_\mu = \begin{pmatrix}
		P_1 + i P_2 & -P_3\\ -P_3 & -P_1 + i P_2
	\end{pmatrix}
\ee
where $(\sigma^\mu)^\alpha{}_{\beta}$ are the Pauli matrices and we raise and lower $SU(2)$ indices with the Levi-Civita tensor $\epsilon_{\alpha \beta}$, with the convention that $\epsilon_{+-} = \epsilon^{+-} = 1$, as $\chi_\alpha = \epsilon_{\alpha \beta} \chi^\beta$ and $\chi^\alpha = \chi_\beta \epsilon^{\beta \alpha}$. The $U(1)_R$ $R$-symmetry rotates $Q_\alpha$ (resp. $\ol{Q}_\alpha$) with weight $-1$ (resp. $1$). We note that $Q_{HT} := \ol{Q}_+$ is a square-zero supercharge and, moreover, the momenta $P_{\ol{z}} = \frac{1}{2}(P_1 + i P_2) = \{Q_{HT}, \frac{1}{2}Q_{+}\}$ and $P_t = P_3 = \{Q_{HT}, -Q_{-}\}$ belong to the image of the twisting supercharge $Q$. In particular, we can consider the $Q_{HT}$ twist (i.e. $Q_{HT}$-cohomology) of our 3d $\CN=2$ theory and the resulting twisted theory behaves holomorphically in the $x^1$-$x^2$ plane and topologically along the remaining $x^3$ direction; the twist by $Q_{HT}$ is correspondingly called the holomorphic-topological ($HT$) twist. 

Due to the holomorphic-topological nature of the twisted theory, we can expect to naturally put it on a 3-dimensional manifold $\CM$ that locally looks like $\C_{z,\ol{z}} \times \R_t$ with transition functions $(z,\bar{z}, t) \to (w(z), \ol{w}(\ol{z}), s(t, z, \bar{z}))$ compatible with the complex structure. Manifolds with this structure, compatible with the HT twist, are said to admit a transverse holomorphic foliation (THF) \cite{ACMV}. To perform the twist, we need to use the $U(1)_R$ $R$-symmetry to define a twisted spin $\textrm{Spin}(2)'$ rotating $\C_{z,\ol{z}}$ under which the supercharge $Q_{HT}$ is a scalar; this choice is called a twisting homomorphism, and we take $J = \frac{1}{2}R - J_3$. With this choice, the various fields of the $\CN=2$ theory will have their transformation properties modified to reflect their twisted spin; for example, a scalar of $R$-charge $r$ will become a section of the $r/2$-th root of the canonical bundle $K^{r/2}$, i.e. it naturally comes with a factor $(\diff z) ^{r/2}$.

We will focus on $\CN=2$ theories of gauged chiral multiplets, often coupled with a superpotential. The data of such a theory is: a choice of compact gauge group $G_c$ (with complexification $G$) and unitary (complex) representation $R$ of $G_c$ ($G$); a choice of Chern-Simons levels $k \in H^4(BG_c)$, whose precise quantization depends on the choice of $R$ (roughly, $k$ is a collection of integers or half-integers for each simple factor of $G_c$); and a choice of gauge-invariant holomorphic function $W: R \to \C$, the superpotential. We will further require that the theory realizes the $U(1)_R$ $R$-symmetry of the $\CN=2$ algebra. This implies that there is an $R$-charge assignment of the chiral multiplets $R = \bigoplus_r R_r$, where $R_r$ is the subrepresentation chiral multiplets of $R$-charge $r \in \Z$, so that $W$ has $R$-charge $2$.

The work \cite{ACMV} provided a remarkably compact description of the $HT$ twist of this class of $\CN=2$ gauge theories in the Batalin-Vilkovisky (BV) formalism; see also the more recent work \cite{CDG}. To write down the theory, we need to introduce some notation. Let $\Omega$ denote the differential graded algebra of forms on $\CM$. The THF on $\CM$ implies that there is a natural subalgebra of differential forms proportional to $\diff z$; we denote by $\BOmega$ the quotient of all differential forms by those proportional to $\diff z$. The THF on $\CM$ also gives us a distinguished line bundle $K$, the canonical bundle on the complex plane $\C$ (or, more generally, the complex leaves of the THF), and we denote by $\BOmega^{(j)}$ the quotient of all differential forms with values in $K^j$ by those forms proportional to $\diff z$. There is a natural product $\BOmega^{(j)} \times \BOmega^{(j')} \to \BOmega^{(j+j')}$ induced by the wedge product of forms as well as a natural integration map $\int: \BOmega^{(1)} \to \C$. In local coordinates, sections of $\BOmega^{(j)}$ look like elements of $C^\infty(\R^3)[\diff t, \diff \ol{z}] \diff z^{j}$, where $\diff t$ and $\diff \ol{z}$ are treated as fermionic variables and $\diff z$ a bosonic variable. We denote by $\diff'$ the fermionic differential on $\BOmega^{(j)}$ induced by the de Rham differential; in local coordinates it takes the form $\diff' = \pd_t \diff t + \pd_{\ol{z}} \diff \ol{z}$. Similarly, we denote by $\pd$ the natural bosonic derivation mapping $\BOmega^{(j)} \to \BOmega^{(j+1)}$; in local coordinates $\pd = \pd_z \diff z$.

With this notation, we can finally state the model of \cite{ACMV} for the HT twist of the $\CN=2$ gauge theory above. First, a chiral multiplet $\Phi$ of $R$-charge $r$ turns into a bosonic field $\BPhi \in \BOmega^{(r/2)}$ and a fermionic field $\BPsi \in \Pi \BOmega^{(1-r/2)}$, where $\Pi$ denotes a parity shift -- the lowest component of $\BPsi$ is a fermion whereas the lowest component of $\BPhi$ is a boson.%
\footnote{Somewhat more precisely, the fields $\BPhi$ and $\BPsi$ have homogeneous Grassmann parity (bosonic vs fermionic), but their components do not. For example, the 0-form and 2-form parts of $\BPhi$ are bosonic differential forms (the components are bosonic) whereas the 1-form part is a fermionic differential form (the components are fermions).} %
In terms of the components with homogeneous form degree, we can write $\BPhi = \phi + \eta^* + \psi^*$ and $\BPsi = \psi + \eta + \phi^*$; $\phi$ and $\psi$ are identified with the complex scalar in the chiral multiplet and a component of the fermion in the anti-chiral multiplet, respectively. The higher form fields are the holomorphic-topological descendants of $\phi$ and $\psi$. For example, the components of $\eta$ are identified with derivatives on the conjugate scalar $\ol{\phi}$. In the same vein, a vector multiplet $V$ turns into an adjoint-valued fermionic field $\BA \in \Pi \BOmega^{(0)} \otimes \fg$ and a coadjoint-valued bosonic field $\BB \in \BOmega^{(1)} \otimes \fg^*$; we expand these fields in components as $\BA = c + A + B^*$ and $\BB = B + A^* + c^*$. In this expansion, the 1-form $A = A_t \diff t + A_{\ol{z}} \diff \ol{z}$ is identified with components of the physical gauge field ($A_t$ is actually complexified by the real scalar $\sigma$), $c$ is the BRST ghost, and $B$ is identified with $F_{zt}$ up to Chern-Simons terms. In pure Chern-Simons theory, or the $HT$ twist of its $\CN=2$ supersymmetric enhancement, $B = B_a T^a$ is identified with $K_{ab} A^b_z$, for $K_{ab}$ the pairing used in the Chern-Simons action, cf. \cite[Section 2.3]{ACMV}. The $HT$-twisted theory is described by the following action:
\be
	S = \int \BB F'(\BA) + \tfrac{k}{4\pi} \Tr(\BA \pd \BA) + \BPsi \diff'_{\BA} \BPhi + \BW
\ee
In this expression, $\diff'_{\BA} = \diff' + \BA$ is the covariant exterior derivative and $F'(\BA) = \diff' \BA + \BA^2$ is its curvature. We have suppressed the natural gauge-invariant pairings of a representation and its dual; $\Tr$ denotes the non-degenerate bilinear form on $\fg$ used for the Chern-Simons terms, e.g. the Killing form with a suitable normalization, with $k$ the (possibly vanishing) level.

In the BV formalism, the equations of motion are encoded in a square-zero supercharge $Q$:
\be
\begin{aligned}
	Q \BA & = F'(\BA) \qquad & Q \BB & = \diff'_{\BA} \BB - \Bmu + \tfrac{k}{2\pi} \pd \BA\\
	Q \BPhi & = \diff'_{\BA} \BPhi \qquad & Q \BPsi & = \diff'_{\BA} \BPsi + \frac{\pd \BW}{\pd \BPhi}
\end{aligned}
\ee
In the present context, the supercharge $Q$ is a combination of two supercharges: the BV/BRST supercharge of the physical theory $Q_{BV}$ as well the $HT$ supercharge $Q_{HT}$. Finally, we note that this theory has a natural $\Z$ grading, traditionally called ghost number, but actually contains a contribution from the $R$ charge $r$ and form degree; the vector multiplet fields $\BA$ and $\BB$ are degree $1$ and $0$, respectively, while the fields $\BPhi$ and $\BPsi$ in a chiral multiplet of $R$-charge $r$ are in degree $r$ and $1-r$, respectively. The ghost numbers of the component fields are uniquely determined by the convention the differential forms $\diff t, \diff \ol{z}$ are given degree $1$ whereas $\diff z$ is given degree $0$. We note that this grading is cohomological, i.e. the supercharge $Q$ has ghost number $1$.

Perturbative local operators in the $HT$ twist can be realized by $Q$-cohomology classes of $G$-invariant operators built from the lowest components $\phi, \psi, \pd c, B$ of the fundamental fields and their $\pd$ derivatives.%
\footnote{The fact that one should consider $\pd c$ and not $c$ itself is due to the fact that its covariant derivative $\CD_z c$ is cohomologous to a gaugino $\lambda_-$; local operators should not include an anti-derivative of this gaugino, cf. \cite[Section 3.2]{CDG}. Although this seems ad hoc, it turns out to be quite natural from the perspective of taking gauge invariants in a derived fashion, i.e. in a way compatible with taking $Q$ cohomology -- unlike taking gauge invariants for infinitesimal gauge transformations, i.e. considering Lie algebra cohomology, taking gauge invariants for constant, finite $G$ gauge transformations is an exact functor and, in particular, should not be done with a ghost. The group of all gauge transformations $\CG$ takes the form of a semi-direct product of $G$ and a unipotent group; since a unipotent group is equivalent to its (nilpotent) Lie algebra, taking derived $\CG$ invariants involves taking $G$ invariants by hand and using ghosts for the rest.} %
In order to access the full, non-perturbative algebra of local operators it is useful to use a state-operator correspondence to relate these operators to states of the theory on a sphere $\PP^1$. The space of such states can be obtained from the above twisted description via geometric quantization of the moduli space of solutions to the equations of motion on $\PP^1$. 

Consider a spacetime of the form $\PP^1 \times \R_t$. The ghost number $0$ part of the equation of motion $F'(\BA) = 0$ reads $\pd_t A_{\ol{z}} = 0$ in a gauge where $A_t = 0$. Thus, the covariant derivative $\CD_{\ol{z}} = \pd_{\ol{z}} + A_{\ol{z}}$ is a connection on a holomorphic $G$ bundle $E$ on $\PP^1$ and this bundle is locally constant in  $t$. One salient feature of this analysis is that the moduli space of solutions naturally fibers over the moduli space $\textrm{Bun}_G(\PP^1)$ of holomorphic $G$ bundles on $\PP^1$ -- many interesting aspects of local operators are encoded in this simple fact.

It is worth noting that it is common to use a slightly different moduli space of $G$ bundles. Instead of considering a finite sphere $\PP^1$, it is more convenient to work with the ``formal bubble'' or ``raviolo'' $\mathbb{B}$. Roughly speaking, the formal bubble is the non-Hausdorf space obtained by gluing two infinitesimally small disks (alias ``formal disks'') away from the origin. The formal bubble naturally arises upon consideration of the state-operator correspondence in any mixed holomorphic-topological quantum field theory: states on the formal bubble $\mathbb{B}$ are expected to be identified with local operators at the center of the bubble. In essence, any sufficiently small neighborhood in a THF 3-manifold takes the form of a small cylinder $D \times I$; the holomorphic-topological nature of the QFT implies that field configurations away from operator insertions are holomorphic on $D$ and constant along $I$. In particular, the size of the interval $I$ is inconsequential and the data of the field configuration is encapsulated by data on $D$. In the presence of a local operator, however, this holomorphic data on $D$ can jump. This jump cannot be arbitrary and should reflect the local nature of the operator: the holomorphic data should be equivalent away from the insertion point of the local operator. Putting this together, field configurations on the boundary of this infinitesimal cylinder can be identified with two copies of the necessary holomorphic data on $D$ and an isomorphism of this data away from the origin. See e.g. the discussion just before Section 3.1 of \cite{Zeng:2021zef} for more details.

The formal bubble can be constructed in an analogous manner to $\PP^1$ by gluing two open sets to one another. One first replaces the complex plane $\C$, with its algebra of functions given by polynomials in a single variable $\C[z]$, by the formal disk $\mathbb{D}$, with algebra of functions given by formal series in a single variable $\CO = \C[\![z]\!]$; one should think of $\mathbb{D}$ as an algebro-geometric avatar of the above infinitesimally small disk. Just as $\PP^1$ is realized by gluing two copies of $\C$ over $\C^\times$, the bubble $\mathbb{B}$ is realized by gluing two copies of the formal disk $\mathbb{D}$ over the ``formal punctured disk'' $\overset{\circ}{\mathbb{D}}$. The algebra of functions on the formal punctured disk is given by formal Laurent series in a single variable $\CK = \C(\!(z)\!)$, cf. functions on $\C^\times$ are Laurent polynomials. Aside from working with series over polynomials, the main difference between the formal bubble is that the transition function relates the two formal disks $\mathbb{D}_z$ and $\mathbb{D}_w$ via $w = z$, rather than the more familiar $w = 1/z$ used for $\PP^1$.

The convenience of working with the formal bubble $\mathbb{B}$ is to provide a rather concrete realization of the moduli space of holomorphic bundles as a particularly nice coset. If we first trivialize the bundle on the two formal disks, it suffices to give the transition function, a gauge transformation on the formal punctured disk $\overset{\circ}{\mathbb{D}}$; the moduli space of holomorphic bundles is then realized as a quotient by changes of trivialization (gauge transformations on the two patches or formal disks). The group of gauge transformations on $\overset{\circ}{\mathbb{D}}$ will be denoted $G_\CK$; when $G$ is a matrix group, elements of $G_\CK$ are simply matrices with Laurent series entries. Similarly, the group of gauge transformations on $\mathbb{D}$ will be denoted $G_\CO$; this is the group with Taylor series entries. Together, we see that the moduli space of $G$ bundles on $\mathbb{B}$ is a double coset 
\be
	\textrm{Bun}_G(\mathbb{B}) = G_\CO \backslash G_\CK / G_\CO = G_\CO \backslash \textrm{Gr}_G
\ee
where the single coset $\textrm{Gr}_G = G_\CK / G_\CO$ is known as the affine Grassmannian. The two factors of $G_\CO$ act on the left and right of $G_\CK$: $(h_1, h_2) \cdot g = h_1 g h_2^{-1}$.

Practically speaking, this moduli space is much easier to manage than its $\PP^1$ counterpart simply due to the fact that formal series are easier to handle than Laurent polynomials. For example, ensuring that a matrix of Laurent polynomials is invertible is much harder than ensuring a matrix of Laurent series is invertible -- in the example of $G = \C^\times$, the group $G[z,z^{-1}]$ only contains the elements $z^n$, whereas $G_\CK$ contains all nonzero Laurent series. Geometrically, this difficult arises from the fact that the transition function defining the bundle must be invertible everywhere except at $0$ and $\infty$; in a sense, the formal bubble $\mathbb{B}$ only contains those points, i.e. we don't have to worry about singularities at finite values of $z$, and so the choice of transition function is much less constrained.

\subsubsection{Abelian monopoles}

Consider the case $G_c = U(1), G = \C^\times$; the inequivalent holomorphic $\C^\times$ bundles on $\mathbb{B}$ are labeled by integers $\m \in \Z$. For any non-zero Laurent series $g(z) = a z^\m + ... \in \C^\times_\CK$ there is an invertible Taylor series of the form $h(z) = a + ... \in \C^\times_\CO$ (invertible implies it has a non-zero constant term) so that $g(z) h(z)^{-1} = z^\m$.%
\footnote{If we were to treat the above double coset properly, we should keep track of the fact that there are non-trivial gauge transformations that preserve this transition function -- since the group is abelian, this transition function is preserved if we perform the same gauge transformation on both disks $(h,h) \cdot g = h g h^{-1} = g$.} 
In particular, the moduli space of solutions to the above equations of motion has disconnected components $\textrm{EOM} = \bigoplus \textrm{EOM}_\m$, where $\textrm{EOM}_\m$ denotes the component of the space of solutions for which the underlying gauge bundle has transition function $z^\m$. The states arising from $\textrm{EOM}_\m$ correspond to operators with magnetic charge (or, better, monopole number) $\m$. The integer $\m$ is naturally identified with a cocharacter of $\C^\times$.

We note that from the perspective of $\PP^1$, the above bundles are identified with the line bundles $\CO(\m)$. This line bundle is characterized by the requirement that sections $\sigma$ transform as $\sigma \to \lambda^\m \sigma$ under the homogeneous scaling of projective coordinates $z^\alpha \to \lambda z^\alpha$. In terms of the affine coordinates $z, w = 1/z$, this corresponds to saying a section of $\CO(\m)$ transforms as $\sigma_z(z) \to \sigma_w(w) = w^\m \sigma_z(1/w)$ under the change of coordinates $z \to w$. The line bundle $\CO(m)$ over $\PP^1$ is the holomorphic avatar of the famous (charge $\m$) monopole bundle over $S^2$.

For more general abelian gauge group $T_c = U(1)^r$ the story is nearly identical; the moduli space of equations of motion has disconnected components labeled by $r$-tuples of integers which give rise to states of the associated monopole number. Recent work of Zeng \cite{Zeng:2021zef} described an explicit geometric quantization of the moduli space $\textrm{EOM}$ on $\PP^1$ for the HT twist of abelian $\CN=2$ theories, thereby realizing (the vector space of) local operators in these twisted theories.

Before moving on, we note that the collision of monopole operators is inherited from the group structure on $G_\CK$. We consider two bubbles $\mathbb{B}$, corresponding to two monopole operators, and stack them on top of one another; the moduli space of bundles on the double-bubble is then the double coset
\be
	G_\CO \backslash G_\CK \times_{G_\CO} G_\CK/G_\CO
\ee
where $G_\CK \times_{G_\CO} G_\CK$ denotes equivalence classes of pairs of gauge transformations $[g_1, g_2]$ up to the action of $G_\CO$, i.e. $[g_1, g_2] \sim [g_1 h^{-1}, h g_2]$; this equivalence corresponds to gauge transformations on the shared disk. We can then form the following convolution diagram:
\be
	\textrm{Bun}_G(\mathbb{B}) \times \textrm{Bun}_G(\mathbb{B}) \overset{\pi_1 \times \pi_2}{\longleftarrow} G_\CO \backslash G_\CK \times_{G_\CO} G_\CK/G_\CO \overset{\mu}{\longrightarrow} \textrm{Bun}_G(\mathbb{B})
\ee
where the $\pi_i$ are (the maps on the quotient coming from) the natural projection maps from $G_\CK \times_{G_\CO} G_\CK$ to $G_\CK/G_\CO$ and $G_\CO \backslash G_\CK$ and $\mu$ is (the map on the quotient coming from) the natural multiplication map $[g_1, g_2] \to g_1 g_2$.

An important consequence of the above convolution structure is that monopole operators act by modifying the gauge bundle at the point of the monopole operator. For the simplest monopole operators, this corresponds physically to performing a large gauge transformation (mathematically this process is called a Hecke modification). For example, we could perform a meromorphic gauge transformation of the form $g = z^\m$; such a gauge transformation maps a section $s$ of $\CO(n)$ to a section $g s$ of $\CO(n+\m)$, defined in terms of affine coordinates as $(g s)_z (z) = z^\m s_z(z)$.%
\footnote{Strictly speaking, this gauge transformation doesn't give us a map on sections. Indeed, generic smooth sections of $\CO(n)$ are mapped to singular sections of $\CO(n+\m)$: when $\m > 0$ (resp. $\m < 0$), the section $\sigma$ must vanish to order $\m$ at $w = 0$ (resp. $z = 0$) for $g \sigma$ to be smooth. A more precise statement is that $g$ defines a \emph{correspondence} between sections of $\CO(n)$ and sections of $\CO(n+\m)$. See, e.g., \cite{BDGHK16} for a discussion of correspondences in the context of monopole operators.} %

\subsubsection{Non-abelian monopoles}
The story for general gauge group $G_c$ is somewhat more complicated. We choose a maximal torus $T_c \subseteq G_c$. Any holomorphic $G$ bundle is equivalent to a $T = (T_c)_\C$ bundle of the above form. (We can always perform a $G$ gauge transformations on the two disks to bring the transition function to the chosen maximal torus.) In particular, for every magnetic charge/cocharacter $\m \in \textrm{Hom}(\C^\times, T) \cong \Z^r$, where $r$ is the rank of $G$, we get a holomorphic $G$ bundle whose transition function is again $g = z^\m$.%
\footnote{If $G_c$ is a matrix group and $T$ the diagonal torus, $z^\m$ simply denotes the diagonal matrix with entries $z^{\m_i}$. More generally, it denotes the composition of the $\C^\times$-valued large gauge transformation described above with the cocharacter $m: \C^\times \to T$ to get a $T$-valued large gauge transformation.} %
Two such cocharacters $\m, \m'$ yield equivalent $G$ bundles when $\m$ and $\m'$ differ by the action of the Weyl group. If we choose a collection of positive roots, we can therefore label such bundles by a dominant cocharacter -- one that pairs non-negatively with all positive roots. Passing from the magnetic charge $m$ to the monopole number $\gamma \in \pi_1(G_c)$ corresponds to taking its image under the quotient map $\Lambda^\vee_{\rm weight} \to \Lambda^\vee_{\rm weight}/ \Lambda^\vee_{\rm root} \cong \pi_1(G_c)$.

To orient readers, we note that, for $G_c = U(r)$ and $T_c = U(1)^r$ the diagonal torus, cocharacters of $T$ can be identified with $r$-tuples of integers $\m = (\m_1, ..., \m_r)$, dominant cocharacters have $\m_i \geq \m_j$ if $j \geq i$ (for the standard choice of positive roots), and the monopole number is simply the sum $\m_1 + \ldots + \m_r \in \Z$. Similarly, for $SU(r)$ (with its dialgonal torus) cocharacters are identified with $r$-tuples satisfying $\m_1 + ... + \m_r = 0$ and for $PSU(r) \cong SU(r)/\Z_r$ (with its diagonal torus) are identified with $r$-tuples $(\m_1, ..., \m_r)$ modulo shifts by $(1,...,1)$. Monopole number is trivial for $SU(r)$ since it is simply connected, whereas monopole number for $PSU(r)$ is simply $\m_1 + \ldots + \m_r \mod r$.

A description of holomorphic $G$ bundles in terms of dominant cocharacters as a discrete set of points is inadequate to capture the intricacies of non-abelian monopoles; it is possible to describe a sequence of holomorphic $G$ bundles with a given magnetic charge $\m$ (a dominant cocharacter) such that the limit bundle has a magnetic charge $\m'$ different from $\m$, a phenomenon known as monopole bubbling. As an example of this phenomenon, consider the $GL(2,\C)$ bundle with transition function $g(z) = z^{(2,0)}$ corresponding to a monopole operator with magnetic charge $\m = (2,0)$, i.e. $g$ is the diagonal $2\times2$ matrix with entries $z^2$ and $1$. We can pre- and post-compose with any non-singular gauge transformation to arrive at an equivalent bundle; for example, we can consider the family of holomorphic $GL(2)$ bundles given by the transition function
\be
	\begin{pmatrix}
		z^2 & a z \\ 0 & 1
	\end{pmatrix} = \begin{pmatrix}
		1 & a z\\ 0 & 1
	\end{pmatrix}\begin{pmatrix}
		z^2 & 0\\ 0 & 1
	\end{pmatrix}
\ee
which simply differs form the above bundle by a gauge transformation on the top disk. If we take the limit $a \to \infty$, we find the gauge bundle with magnetic charge $\m' = (1,1)$: away from $a = 0$ we can perform the following gauge transformation on the lower disk
\be
	\begin{pmatrix}
		z^2 & a z\\ 0 & 1
	\end{pmatrix}\begin{pmatrix}
		1 & -a\\ (1-z)/a & z
	\end{pmatrix} = \begin{pmatrix}
		z & 0 \\ (1-z)/a & z
	\end{pmatrix}
\ee
from which the $a \to \infty$ limit is straightforward. In particular, we arrive at a gauge bundle with magnetic charge $\m' = (1,1)$. More generally, the magnetic charge $\m'$ is always dominated by $\m$ (i.e. $\m - \m'$ is dominant) and projects to the same monopole number $\gamma = [\m] = [\m']$. To contrast this with the abelian case above, the moduli space of holomorphic $G$ bundles for general $G$ has disconnected components labeled by $\pi_1(G)$ -- monopole number, and not magnetic charge, is a good quantum number for local operators in 3d.

One of the primary obstacles to the study of monopole operators in $HT$-twisted $\CN=2$ theories with non-abelian gauge groups is having to contend with the highly non-trivial structure of the moduli space of bundles. A related problem arises in the topological $A$-twist of standard $\CN=4$ theories, but the increased amount control offered by the topological twist makes it possible to progress -- the full, non-perturbative algebra of local operators in these theories was be described explicitly by Braverman-Finkelberg-Nakajima \cite{Nak, BFN}; see \cite{BDG} for a physical analysis of this algebra, related to the work of Braverman-Finkelberg-Nakajima via fixed-point localization in equivariant cohomology.%
\footnote{In reality, it is not the moduli space of bundles $G_\CO\backslash \textrm{Gr}_G$ itself that causes the difficulty. Rather, it is the fact that there is generally very little control over the moduli space $\textrm{EOM}$ as one moves around in the moduli space of bundles. The fact that the moduli space of bundles is particularly simple for abelian $G$ offers one way to understand the moduli space $\textrm{EOM}$ in those settings. The fact that one can handle non-abelian gauge groups for the topological $A$-twist is entirely due to the relative simplicity of $\textrm{EOM}$.} %

\subsection{Boundary monopoles}
\label{sec:bdymonopoles}

Although monopole operators in the bulk of a 3d gauge theory are themselves quite fascinating, we will be more interested in their appearance in the algebra of local operators on boundary conditions of these gauge theories. In 3d $\CN=2$ gauge theories, the BPS equations $F = \star D\sigma$ and $D \star \sigma = 0$ for the gauge field strength $F$ and real scalar $\sigma$ admit non-trivial, monopole solutions when the vector multiplets are given half-BPS Dirichlet boundary conditions; see e.g. Section 3.4 of \cite{DGP} for more details. Local operators on this Dirichlet boundary condition can be obtained in a similar fashion to bulk local operators; we can realize them via a state-operator correspondence as states on a hemisphere attached to the boundary. Again, this space of states can be determined via geometric quantization of a suitable space of equations of motion. We note that the chiral algebras studied in \cite{ADLslab}, realized as algebras of local operators on an interval compactification of 3d $\CN=2$ theories rather than on a boundary, share many of the features we describe below.

It is convenient to work with an infinitesimal version of the hemisphere, i.e. a formal disk $\mathbb{D}$ with the boundary conditions imposed on its boundary formal punctured disk $\overset{\circ}{\mathbb{D}}$; the Dirichlet boundary conditions require that the holomorphic gauge bundle is equipped with a trivialization on the boundary. Following the argument above, we see that the moduli space of solutions to the equations of motion naturally fibers over the affine Grassmannian $\textrm{Gr}_G$, which parameterizes such bundles. In particular, we again see that these equations of motion have disconnected components labeled by monopole number $\pi_1(G)$ -- this is again a good quantum number for boundary local operators. The fact that we arrive at the affine Grassmannian, rather than its quotient by $G_\CO$, is due to the fact that we are not required to impose gauge invariance of local operators on Dirichlet boundary conditions.

The fact that the space of solutions to the equations of motion fibers over the affine Grassmannian $\textrm{Gr}_G$ can be a powerful conceptual tool, but the full, non-perturbative algebra of local operators can still be quite difficult to describe. The work \cite{CDG} provided a conjectural description of the algebra of local operators as follows; see Section 7 of loc. cit. for more details. Suppose the algebra of local operators in the absence of gauge fields realizes some vertex algebra $\CV$ (it need not have a holomorphic stress tensor); $\CV$ necessarily admits an action of $G_\CO$. We can then use this action to construct an interesting vector bundle $\CV_{\textrm{Gr}_G}$ over the affine Grassmannian as the product $\CV_{\textrm{Gr}_G} = G_\CK \times_{G_\CO} \CV$; for a given holomorphic $G$ bundle $P$, the fiber $\CV_P$ over $P$ should be thought of as $\CV$ coupled to the bundle $P$. The algebra of local operators on this Dirichlet boundary condition is then conjectured to be the Dolbeault homology of $\textrm{Gr}_G$ with coefficients of this vector bundle twisted by a certain power of the determinant line bundle $\CL$:
\be
\label{eq:dirichletBC}
	H_{\bullet, \ol{\pd}}(\textrm{Gr}_G, \CV_{\textrm{Gr}_G} \otimes \CL^{-\kappa}) = \bigoplus_n H_{n, \ol{\pd}}(\textrm{Gr}_G, \CV_{\textrm{Gr}_G} \otimes \CL^{-\kappa})
\ee
where, for a complex manifold $X$, $H_{n, \ol{\pd}}(X, \CE)$ is defined to be the linear dual of the Dolbeault cohomology valued in the dual bundle $H_{n, \ol{\pd}}(X, \CE) := H^n_{\ol{\pd}}(X, \CE^*)^*$, cf. Eq. (7.13) of loc. cit. See, e.g., \cite[Section 1.5]{Zhu} for a detailed description of the line bundle $\CL \to \textrm{Gr}_G$. The power $\kappa$ denotes the effective Chern-Simons level, cf. \cite[Eq. 3.32]{DGP}, and encodes the anomaly for the $G$ flavor symmetry on the boundary. Practically speaking, the effect of the correction $\CL^{-\kappa}$ is to give magnetically charged operators (boundary monopoles) a non-trivial electric charge in the presence of a Chern-Simons term. It is important to note that Eq. \eqref{eq:dirichletBC} is merely a description of the vector space of local operators. Equipping this vector space with the structure of a vertex algebra is far from trivial; this is sketched in \cite[Section 7.2]{CDG} in the absence of matter fields.

\subsubsection{Boundary monopoles in abelian gauge theories and spectral flow}
\label{sec:abelian3dbdy}

The relative simplicity of the affine Grassmannian $\textrm{Gr}_T$ for abelian gauge group $T$ once again implies that the description of boundary local operators in Eq. \eqref{eq:dirichletBC} can be made fairly concrete. Recall that $\textrm{Gr}_T$ is essentially the lattice $\Z^r$ associated to the holomorphic $T = (\C^\times)^r$ bundles with transition function $z^\m$; the vector space $H_{\bullet, \ol{\pd}}(\textrm{Gr}_T, \CV_{\textrm{Gr}_T} \otimes \CL^{-\kappa})$ is thus graded by magnetic charge/monopole number $\Z^r$.

Local operators with vanishing magnetic charge are simply the perturbative local operators $\CV$ built from the fundamental fields; since magnetic charge is additive, the full, non-perturbative algebra furnishes a module for this perturbative subalgebra. By construction, this subalgebra has abelian currents $J_a$, $a = 1, ..., r$ that generate the action of the $T$ flavor symmetry on the boundary. The statement this flavor symmetry has an anomaly $\kappa$ translates to the OPEs
\be
	J_a(z) J_b(w) \sim \frac{\kappa_{ab}}{(z-w)^2}.
\ee
The OPEs of these currents with a perturbative local operator $\CO$ built from the matter fields simply measures it charge $q_a$ under $T$:%
\footnote{Strictly speaking, this is the OPE of $J_a$ with a so-called primary operator. A general operator can be realized as a (linear combination of) derivatives of primary operators.} %
\be
	J_a(z) \CO(w) \sim \frac{q_a}{z-w} \CO(w).
\ee

Given the OPEs with $J_a$, we can immediately write down the result of coupling to the gauge bundle with magnetic charge $\m \in \Z^r$ by performing the large gauge transformation $z^\m$. Let $\m^a \in \Z$ denote the components of the magnetic charge. If a local operator $\CO$ has charges $q_a$ then it transforms as $\CO \to z^{\langle q,\m\rangle}\CO$, where $\langle q,\m\rangle = \sum q_a \m^a$ is the natural pairing of the weight $q \in \textrm{Hom}(T, \C^\times)$ and cocharacter $\m \in \textrm{Hom}(\C^\times, T)$. The currents $J_a$, on the other hand, transform as $J_a \to J_a - \frac{\kappa_{ab}\m^b}{z}$; if $\kappa_{ab}$ is non-degenerate, one could say that $\kappa^{ab}J_b$ transforms as a connection. Indeed, this is no mistake: the current $J_a$ can be identified with the boundary value of the field $B_a$ appearing in Section \ref{sec:N=2HT}, which itself is identified with component $A^a_z$ of the gauge field as $B_a = \kappa_{ab} A^b_z$.

It is important to note that the transformations just described are automorphisms $\sigma_\m$ of the (mode algebra of the) perturbative subalgebra known as spectral flow automorphisms; for every abelian current $J$, there is a lattice of such automorphisms. More generally, if $B$ is an abelian current, then there is a family of automorphism of its mode algebra that sends $B \to B + \frac{\lambda}{z}$; if this is a current subalgebra of a larger vertex algebra $\CV$, it need not lift to an automorphism of all of $\CV$, cf. Section 2.4 of \cite{BCDN23}. For example, if $\CO \in \CV$ is a current algebra primary of weight $q$ then we need $\lambda q \in \Z$. The inverse of $\sigma_\m$ is simply $\sigma_{-\m}$. Given the automorphism $\sigma_\m$ and a module $M$, we can construct a new module $\sigma_m(M)$, called the spectral flow of $M$, by pre-composing with the automorphism: the action of an operator $\CO$ on the module $\sigma_m(M)$ is given by the action of $\sigma_{-\m}(\CO)$ on $M$:
\be
	\CO \sigma_\m(\ket{\psi}) = \sigma_{\m}(\sigma_{-\m}(\CO)\ket{\psi})
\ee
In the present situation, we are interested in the spectral flows of the vacuum module $\CV$; we claim that, as a module for the perturbative algebra $\CV$, the full, non-perturbative algebra of local operators is simply the direct sum over all such spectral flows: $\bigoplus_{\m \in \Z^r} \sigma_\m(\CV)$. It is worth noting that the notion of spectral flow and spectral flow modules is totally independent of the presence of a 3d bulk; for example, (the symmetry algebra of) a 2d free boson $\varphi(z,\ol{z})$ has spectral flow automorphisms and the resulting modules are generated by the usual vertex operators $\norm{e^{\lambda \varphi}}$. When a vertex algebra $\CV$ with spectral flow automorphisms $\sigma_n$ is realized as the boundary vertex algebra for a 3d TQFT, the spectral flows of the vacuum $\sigma_n(\CV)$ are identified with vortex line operators and generator of a 1-form symmetry of the 3d TQFT, cf. Physics Proposition 6.2 of loc. cit.

To illustrate how this works in practice, consider the simplest case of a single abelian current $J$ at level $\kappa \in \Z_{>0}$ without any matter fields. The vacuum module of this current algebra is generated by the vacuum vector $\ket{0}$ on which the modes $J_{n}$ of $J = \sum J_n z^{-n-1}$ act as
\be
	J_n \ket{0} = \begin{cases} 0 & n \geq 0\\ J_n \ket{0} & n < 0 \end{cases}
\ee
The vacuum module $\CV$ is then spanned by vectors of the form $J_{n_1} ... J_{n_m} \ket{0}$ where $n_1 \leq ... \leq n_m < 0$. The spectral flow automorphism $\sigma_\m$ takes the form $\sigma_\m(J)(z) = J(z) - \frac{\kappa m}{z}$ and is simply a shift in the zero mode $J_0$. This implies that the spectral flow of the vacuum module $\sigma_\m(\CV)$ is generated by the spectral flow of the vacuum vector $\sigma_\m(\ket{0})$, on which the modes act as
\be
	J_n \sigma_\m(\ket{0}) = \sigma_\m(\sigma_{-\m}(J_n) \ket{0}) = \begin{cases} 0 & n > 0\\ \kappa m \sigma_\m(\ket{0}) & n = 0\\ \sigma_\m(J_n \ket{0}) & n < 0 \end{cases}
\ee
The spectral flow of the vacuum vector $\sigma_\m(\ket{0})$ corresponds to an operator $V_\m(0)$, the boundary monopole of magnetic charge $\m$; this action of the modes $J_n$ corresponds to the OPE
\be
	J(z) V_\m(w) \sim \frac{\kappa \m}{z-w} V_\m(w).
\ee
As claimed above, the effect of a non-trivial level $\kappa$ is to give the magnetically charged operator $V_\m(w)$ a non-trivial electric charge $\kappa \m$.

The above analysis readily generalizes to more interesting examples. That said, it is important to note that the precise OPEs of the boundary monopole operators is not encoded in the action of the perturbative algebra on the sectors of non-trivial magnetic charge. It is believed that the spectral flow morphisms $\sigma_\m$ are compatible with the fusion of modules for the perturbative algebra $\CV$; given two modules $M_1, M_2$ it is expected that $\sigma_{\m_1}(M_1) \times \sigma_{\m_2}(M_2) \simeq \sigma_{\m_1 + \m_2}(M_1 \times M_2)$. In particular, since $\CV \times \CV = \CV$, the fusion of the modules $\sigma_\m(V)$ respects the grading by magnetic charge $\m$. The direct sum over the $\sigma_\m(\CV)$ is thus a simple current extension of the perturbative algebra $\CV$.

In practice, the precise OPEs of the fields in $\sigma_{\m_1}(\CV)$ and $\sigma_{\m_2}(\CV)$ are determined by a free-field realization of $\CV$. In the above example of the abelian current $J$, we can realize it as $J = \pd \varphi$ for a chiral boson $\varphi(z)$; the boundary monopole operators $V_\m(z)$ are then identified with the vertex operators $\norm{e^{\m \varphi}}$. The full, non-perturbative algebra in this example is thus identified with a lattice VOA. See e.g. \cite{BN22, GN23, BCDN23} for examples of how this process can be implemented in more interesting examples; the work \cite{BN22} studies the algebra of local operators on a Dirichlet boundary condition in the (topological) $B$-twist of 3d $\CN=4$ SQED, whereas the more recent \cite{GN23, BCDN23} apply these same techniques to study yet more involved abelian $\CN=4$ gauge theories. The much earlier work \cite{CR13} studies related simple current extensions from a purely VOA perspective.

Before moving on, we note that the sorts of simple current extensions that arise from spectral flows of the vacuum module have an important physical role to play. General aspects of these simple current extensions were described in \cite{CKM17, CMY22}, which we now sketch. For starters, given any module $M$ of the perturbative algebra we can consider the direct sum $\bigoplus_\m \sigma_\m(M)$. Importantly, not all $\bigoplus_\m \sigma_\m(M)$ will yield a module for the full non-perturbative algebra: the OPE of a monopole of magnetic charge $\m$ and a weight $q$ operator $\CO$ takes the form 
\be
	V_\m \CO \sim (z-w)^{q\m} \sigma_\m(\CO) + \dots
\ee
To ensure we get an honest module for the simple current extension, rather than a twisted module, requires the exponent $q\m$ must be integral for all $\m$, i.e. $q$ is integral. In particular, 
\begin{enumerate}
	\item boundary monopoles will disallow the modules $M$ that do not have integral weights
\end{enumerate}
Moreover, given a perturbative module $M$ that induces a module for the non-perturbative algebra, the map $M \mapsto \bigoplus_\m \sigma_\m(M)$ is not injective: the modules $M$ and $\sigma_{\m'}(M)$ will both yield the same non-perturbative module. If we wish to describe modules for the non-perturbative algebra in terms of modules for the perturbative algebra, we see that 
\begin{enumerate}
	\item[2.] boundary monopoles identify perturbative modules that differ by spectral flow
\end{enumerate}
One often phrases this phenomenon physically as ``Wilson line are being screened by gauge vortex lines.''

To make this more explicit, consider again the case of $U(1)_k$ Chern-Simons theory. The perturbative algebra is simply a Heisenberg algebra at level $k$; the full non-perturbative algebra obtained by extending by spectral flows of the vacuum can be identified with the lattice VOA based on $\sqrt{k} \Z$. There is a Fock module $F_q$ for the perturbative Heisenberg algebra for each infinitesimal weight $q \in \C$, viewed as a Wilson line for a representation of weight $q$; spectral flow acts by shifting the weight $\sigma_\m(F_q) = F_{q + k\m}$. A straightforward computation shows that $\bigoplus \sigma_\m(F_q)$ defines a module for the lattice VOA so long as $q \in \Z$, i.e. we can only consider Wilson lines of integer charge. Finally, since $F_q$ and $F_{q+k\m}$ induce the same module for the lattice VOA, we see that gauge vortex lines screen Wilson lines with charge a multiple of $k$, i.e. that modules for the lattice VOA are labeled by integers mod $k$.

\subsubsection{Boundary monopoles in non-abelian gauge theories}

Perhaps unsurprisingly, boundary monopoles in non-abelian gauge theories are notably more difficult than the case of abelian gauge theories. This increase in difficulty is directly tied to the non-trivial geometry and topology of the affine Grassmannian for non-abelian groups.

Even in the case of a lone current algebra, i.e. in the absence of the matter VOA $\CV$, the story of boundary monopoles is not well understood. To illustrate the difficulty, consider for simplicity $SU(2)$ Chern-Simons theory at level $\kappa$, a positive integer. Perturbatively, the algebra of boundary local operators on the above Dirichlet boundary condition is generated by three holomorphic currents $J_H, J_E, J_F$ (labeled by the three Chevalley generators of $\fsl(2)$ denoted by $H, E, F$),  with singular OPEs
\be
\begin{aligned}
	J_H(z) J_H(w) & \sim \frac{2\kappa}{(z-w)^2} & \qquad J_E(z) J_F(w) & \sim \frac{\kappa}{(z-w)^2} + \frac{J_H(w)}{z-w}\\
	J_H(z) J_E(w) & \sim \frac{2J_E(w)}{z-w} & \qquad J_H(z) J_F(w) & \sim \frac{-2J_F(w)}{z-w}.\\
\end{aligned}
\ee
On the other hand, the full, non-perturbative boundary algebra is expected to be a WZW current algebra \cite{DGP, CDG}; this is a simple VOA obtained by removing singular vectors, essentially $\norm{J_E{}^{\kappa+1}}$ and its descendants, from the perturbative current algebra. We immediately see a dramatic difference from the abelian setting: the full, non-perturbative algebra is a quotient of the perturbative algebra, rather than the simple-current extension appearing in the abelian setting. More generally, when the gauge group $G$ isn't simply connected, the non-perturbative algebra is an extension of the simple quotient by simple currents furnishing the fundamental group $\pi_1(G)$ -- the simple-current extension is thus graded by the (abelian) group $\pi_1(G)$ capturing the decomposition into sectors of a given monopole number.

With this said, we note that there are still spectral flow automorphisms $\sigma_\m$ which can be thought of as considering perturbations around the holomorphic $SL(2)$ bundle with transition function $z^{(\m,-\m)}$; equivalently, we can think of this as performing a large gauge transformation on the perturbative algebra by the diagonal matrix
\be
	g(z) = \begin{pmatrix} z^\m & 0\\ 0 & z^{-\m}\end{pmatrix}.
\ee
We again identify the $J_a$ with the boundary values of the gauge fields $A^a_z$ via $J_a = \kappa K_{ab} A^b_z$, where $K_{ab} = \Tr(t_a t_b)$ is the Killing form. Thus, the currents transform as
\be
\begin{aligned}
	J_H(z) & \to \sigma_\m(J_H(z)) = J_H(z) - \frac{2\kappa \m}{z}\\
	J_E(z) & \to \sigma_\m(J_E(z)) = z^{-2\m} J_E(z)\\
	J_F(z) & \to \sigma_\m(J_F(z)) = z^{2\m} J_F(z).\\
\end{aligned}
\ee
We note that there is a square root $\sigma_{1/2}$ of the spectral flow automorphism $\sigma_1$. This square root should be thought of as performing a large gauge transformation by the diagonal matrix $z^{(1,0)} \sim z^{(0,-1)}$ in a related $SO(3) \cong PSU(2)$ gauge theory.

Although we will not address the full problem here, we note that there are hints that it is possible to connect the two perspectives: it should be possible to extend the perturbative subalgebra by modules $\CE_\m$ labeled by cocharacters $\m$ together with a differential encoding the geometry of the affine Grassmannian $\textrm{Gr}_G$. The modules $\CE_\m$ should be thought of as contributions from the torus fixed points on $\textrm{Gr}_G$ (the torus action arising from left multiplication on $G_\CK$) with the differential encoding how these fixed points fit into honest homology classes. We are unaware of a description of the $\CE_\m$, but mention that the spectral flows of the vacuum are likely submodules thereof: the index computations of \cite[Section 7.1]{DGP} illustrate that, upon analytic continuation, the characters of the full non-perturbative algebra can be matched with the direct sum of these spectral flows.

The fact that they only match after analytic continuation is an indication that the spectral flow modules are extended by modules whose characters vanish upon analytic continuation. For a simple illustration of this fact, consider the algebra $\C[X]$ of polynomials in a bosonic variable $X$ and the module $M_\m = X^\m \C[X]$. If we count powers of $X$ with a fugacity $q$, the character of $M_\m$ is simply
\be
	\chi(M_\m) = \sum_{i=0}^{\infty} q^{\m+i}.
\ee
So long as $|q| < 1$ this character converges to the rational function $q^\m/(1-q)$; upon analytic continuation to $|q|> 1$ and then expansion as a series in powers of $q$ yields $-1$ times the character of the quotient module $\C(X)/M_\m$:
\be
	\chi(M_\m) \overset{|q|<1}{=} \frac{q^\m}{1-q} \overset{\textrm{analytic cont.}}{\longrightarrow} \frac{-q^{\m-1}}{1-q^{-1}} \overset{|q| > 1}{=} -\sum_{i=0}^{\infty} q^{\m-1-i} = -\chi(\C(X)/M_\m)
\ee
Of course, the modules $M_\m$ and $\C(X)/M_\m$ are very different, but they fit into the following exact sequence of $\C[X]$-modules:
\be
	0 \to M_\m \to \C(X) \to \C(X)/M_\m \to 0
\ee
In particular, we see that the module $M_\m$ is a submodule of a resolution $0 \to M_\m \to \C(X)$ of the module $\C(X)/M_\m$ obtained by analytically continuing its character; the character of $\C(X)$ is the formal delta function $\delta(q) = \sum_{i \in \Z} q^i$ and vanishes upon analytic continuation to $|q|> 1$.

Translating this lesson to the case of interest, we arrive at the expectation that the spectral flows of the vacuum $\sigma_\m(\CV)$ are themselves submodules of the $\CE_\m$ used to resolve the non-perturbative algebra. Even though the $\sigma_\m(\CV)$ do not constitute the entirety of the non-perturbative algebra outside of abelian gauge theories, they are still an important part to understand.

\section{Twistorial Monopole Operators}
\label{sec:celestial}

We now apply the above analysis to the 3d theories arising from compactifying holomorphic BF theory on twistor space $\PT$ as described in \cite{CP22}. See Appendix \ref{app:twistor} for our twistorial conventions. We will be principally interested in the effect of non-perturbative monopole operator insertions on the 3d boundary chiral algebra. We remind the reader that precisely this chiral algebra is the asymptotic symmetry algebra of (conformally) soft modes \cite{Guevara:2021abz}, supported on the celestial sphere at asymptotic null infinity from the perspective of the 4d BF theory \cite{CP22}.

\subsection{Self-dual gauge theory from $\PT$}
\label{sec:SDYM}

We start with a brief reminder of the realization of self-dual 4d gauge theory in terms of a holomorphic field theory on $\PT$; we mostly follow \cite{CP22}. Our goal for this subsection is to relate these self-dual gauge theories to 3d theories reminiscent of the twisted theories described in Section \ref{sec:N=2HT} whereby the celestial chiral algebras of asymptotic symmetries are realized as boundary chiral algebras of the 3d theory, cf. Section 6 of loc. cit.

Consider self-dual gauge theory with gauge Lie algebra $\fg$. This is the 4d theory of a gauge field $A \in \Omega^1(\R^4,\fg)$ and self-dual 2-form $B \in \Omega^2(\R^4, \fg^*)_-$ with a BF action:
\be
	S_{\textrm{4d}} = \int_{\R^4} B F(A)_-
\ee
where $F(A)_-$ is the anti self-dual part of the curvature $F(A)$. The Penrose-Ward correspondence relates this self-dual gauge theory to a 6d field theory on twistor space $\PT$; the fields of the 6d theory include a $(0,1)$ gauge field $\CA \in \Omega^{(0,1)}(\PT, \fg)$ together with a $(3,1)$-form field $\CB \in \Omega^{(3,1)}(\PT,\fg^*)$ with a holomorphic BF action given by
\be
	S_{\textrm{6d}} = \int_\PT \CB F^{(0,2)}(\CA)
\ee
where $F^{(0,2)}(\CA) = \ol{\pd}\CA + \CA^2$ is the $(0,2)$ part of the curvature of $\CA$. When formulated in the BV formalism, we can account for the anti-fields, ghosts, and anti-ghosts by extending $\CA$ and $\CB$ to $\Omega^{(0,\bullet)}(\PT,\fg)[1]$ and $\Omega^{(3,\bullet)}(\PT,\fg^*)[1]$, respectively. We denote the extended fields by the same characters and use the convention that integration is only non-vanishing on top forms; with this choice, the BV action takes the same functional form as above.

\subsubsection{Anomaly on $\PT$}
\label{sec:SDYManomaly}

It is important to note that the above 6d holomorphic theories suffer from perturbative anomalies \cite{Costello:2019jsy, Costello:2021bah}. These anomalies do not say that the 4d self-dual gauge theory is ill-defined. Rather, they imply the Penrose-Ward correspondence fails to hold quantum mechanically. Consequently, we should not expect the 3d theory obtained by reduction from 6d to describe the corresponding 4d self-dual gauge theory beyond tree-level computations.

As shown in \cite{Costello:2021bah}, it is possible to remedy this anomaly via a Green-Schwarz-like mechanism for certain special choices of gauge Lie algebra $\fg$. In particular, when the gauge Lie algebra is $\fsl(2)$, $\fsl(3)$, $\mathfrak{so}(8)$, or an exceptional algebra one can introduce a Kodera-Spencer-like field (and its BV partners) $\eta \in \Omega^{(2,\bullet)}(\PT)$ satisfying $\pd \eta = 0$ with full 6d action
\be
	S_{\textrm{6d}} = \int_\PT \CB F^{(0,2)}(\CA) + \tfrac{1}{2} \ol{\pd}\eta \pd^{-1}\eta + \frac{\lambda_\fg}{4(2\pi i)^{3/2} \sqrt{3}} \eta \Tr(\CA\pd\CA)
\ee
where $\lambda_\fg$ is a $\fg$-dependent constant.%
\footnote{Explicitly, the constant $\lambda_\fg$ satisfies $$\Tr(X^4) = \lambda_\fg^2 \textrm{tr}(X^2)^2$$ for $\Tr$ the trace in the adjoint representation and $\textrm{tr}$ is the trace in the fundamental representation. The existence of a non-zero solution implies the above restriction on the gauge Lie algebra $\fg$.} %
From the 4d perspective, the new field $\eta$ introduces an axionic field with a quartic kinetic term. Since a majority of our computations are at tree-level, we will mostly ignore the contributions from this axionic field -- the classical theory with the Kodera-Spencer field is inconsistent, as the 1-loop anomaly of the holomorphic BF theory is canceled by a \emph{tree-level} anomaly involving $\eta$.

\subsubsection{Reduction to 3d}
\label{sec:SDYM3d}

In order to connect the above 4d and 6d theories to the discussion of Section \ref{sec:monopoles}, we consider a radial slicing on $\R^4\backslash\{0\}$ and compactify the 6d theory on $\PT \backslash \PP^1 \simeq \R^4 \backslash\{0\} \times \PP^1$ along the radial $S^3$ slices. The resulting 3d theory on $\R_{>0} \times \PP^1$ remains holomorphic on $\PP^1$ and becomes topological along $\R_{>0}$; the Kaluza-Klein modes along the $S^3$ organize themselves into representations of the $SU(2)$ isometry of $S^3$. The field content of this mixed holomorphic-topological 3d theory can be determined in many ways, see Section 6 of \cite{CP22} for a slightly different approach to the one presented below. Our approach will be to mimic the algebraic reduction outlined in \cite{GW2018} in the context of higher-dimensional Kac-Moody algebras; the recent paper \cite{Zeng2023} of Zeng describes how this reduction can be done in a smooth, rather than algebraic setting; as usual, the algebraic classes are dense in the smooth classes, cf. Appendix B of loc. cit. We use the notation of Section \ref{sec:N=2HT}.

We start by (locally) placing the theory on the open set $\C \times (\C^2\backslash\{0\}) \subset \C^3$. The ring of functions, i.e. the cohomology of the structure sheaf, on this open set has support in nonzero cohomological degree, reflecting the fact that $\C^2\backslash\{0\}$ is not affine. Indeed, a standard computation shows that
\be
	\C[\C^2\backslash\{0\}] \simeq \overset{\textrm{degree } 0}{\overbrace{\C[v^{\dot{\alpha}}]}} \oplus \overset{\textrm{degree } 1}{\overbrace{(v^{\dot{1}}v^{\dot{2}})^{-1}\C[(v^{\dot{\alpha}})^{-1}]}}
\ee
The fact that the degree 0 part of this ring of functions agrees with the ring on functions on all of $\C^2$ is a manifestation of Hartog's Lemma that states any holomorphic function on $\C^n\backslash\{0\}$ extends to a holomorphic function on $\C^n$, so long as $n>1$. Note that the degree 1 subspace can be identified with the dual space to the degree 0 part using the residue pairing. When we reduce to 3d by compactifying around the $S^3$ inside $\C^2\backslash\{0\} \simeq \R_{>0} \times S^3$, each of these cohomology classes will produce a 3d field. For example, the gauge field $\CA$ reduces to two towers of fields:
\be
	\CA \rightsquigarrow \BA[m_1,m_2] \textrm{ and } \BPhi[m_1,m_2]
\ee
where $m_1,m_2 \geq 0$; the 3d fermionic field $\BA[m_1,m_2]$ corresponds to the coefficient of the degree 0 cohomology class $(v^{\dot{1}})^{m_1} (v^{\dot{2}})^{m_2}$ and the 3d bosonic field $\BPhi[m_1,m_2]$ corresponds to the degree 1 cohomology class $(v^{\dot{1}})^{-(1+m_1)} (v^{\dot{2}})^{-(1+m_2)}$. Importantly, these fields inherit a non-trivial 3d spin from the fact that the coordinates $v^{\dot{\alpha}}$ transform a sections of $\CO(1) \to \PP^1$, i.e. the coordinate $v^{\dot{\alpha}}$ has 3d spin $\tfrac{1}{2}$. Correspondingly, $\BA[\vec{m}] \in \BOmega^{(\tfrac{m_1+m_2}{2})} \otimes \fg$ has spin $\frac{m_1+m_2}{2}$ and $\BPhi[\vec{m}] \in \BOmega^{(-1-\tfrac{m_1+m_2}{2})} \otimes \fg$ has spin $-1-\frac{m_1+m_2}{2}$. In a similar fashion, $\CB$ produces towers $\BB[\vec{m}], \BLambda[\vec{m}]$ (spins $1-\frac{m_1+m_2}{2}$ and $2+\frac{m_1+m_2}{2}$). The KK modes with fixed $m_1+m_2$ transform in the spin $\frac{m_1+m_2}{2}$ representation of the $SU(2)$ rotating the $S^3$ slices.

Passing the action $S_{\textrm{6d}}$ through this compactification, we arrive at a theory of the form described in Section \ref{sec:N=2HT}, albeit with an infinite number of fields coming from the KK modes. In order to write it down concisely, it is convenient to collect the KK modes into generating functions
\be
	\BA(v^{\dot{\alpha}}) = \sum_{m_1,m_2} (v^{\dot{1}})^{m_1} (v^{\dot{2}})^{m_2} \BA[\vec{m}] \qquad \BPhi(v^{\dot{\alpha}}) = \sum_{m_1,m_2} (v^{\dot{1}})^{-1-m_1} (v^{\dot{2}})^{-1-m_2} \BPhi[\vec{m}]
\ee
and similarly for the towers coming from $\CB$, cf. Eq. (6.2.3) of \cite{CP22}. With this notation, we can write the action of the 3d theory in a remarkably compact form:
\be
	S_{\textrm{3d}} = \int \oint \frac{\diff v^{\dot{1}}}{2\pi i} \oint \frac{\diff v^{\dot{2}}}{2\pi i} \bigg(\BB(v^{\dot{\alpha}}) F'(\BA(v^{\dot{\alpha}})) + \BLambda(v^{\dot{\alpha}}) \diff'_{\BA(v^{\dot{\alpha}})} \BPhi(v^{\dot{\alpha}}) \bigg)
\ee
where we view $\BA(v^{\dot{\alpha}})$ as a partial connection for the infinite-dimensional Lie algebra $\fg[\![v^{\dot{\alpha}}]\!]$, with covariant exterior derivative $\diff'_{\BA(v^{\dot{\alpha}})}$ and curvature $F'(\BA(v^{\dot{\alpha}}))$. In terms of the component fields, this action takes the following form:
\be
\label{eq:3dSDYM}
\begin{aligned}
	S_{\textrm{3d}} & = \sum_{m_1, m_2}  \int \BB_a[m_2,m_1] \diff' \BA^a[m_1,m_2] + \BLambda_a[m_2,m_1] \diff' \BPhi^a[m_1,m_2]\\
	& + \sum_{m_1,m_2,n_1,n_2} \tfrac{1}{2}\int f^c{}_{ab} \BB_c[m_2+n_2,m_1+n_1] \BA^a[m_1,m_2] \BA^b[n_1,n_2]\\
	& + \sum_{m_1,m_2,n_1,n_2} \int f^c{}_{ab} \BPhi^b[m_2+n_2,m_1+n_1] \BA^a[m_1,m_2] \BLambda_c[n_1,n_2]\\
\end{aligned}
\ee

\subsubsection{The celestial chiral algebra as a boundary chiral algebra}
With this reduction from 6d to 3d, it is a straightforward task to realize the celestial chiral algebra of asymptotic symmetries as a boundary chiral algebra for the above 3d theory using the tools outlined in Section \ref{sec:monopoles}. This phenomenon is quite general: the work \cite{CP22} says that the celestial chiral algebra of asymptotic symmetries of any twistorial theory can be realized as the boundary chiral algebra of a suitable 3d mixed holomorphic-topological theory. The celestial chiral algebra controlling (tree-level) scattering amplitudes is realized by imposing Dirichlet boundary conditions on the gluon ``vector multiplets'' $(\BA, \BB)$ and imposing Neumann boundary conditions on gluon ``chiral multiplets'' $(\BPhi, \BLambda)$. In analogy with \cite[Section 7.1]{CDG}, the perturbative algebra of local operators is generated by the boundary values of the bosons $B[\vec{m}], \Phi[\vec{m}]$. We denote $B, \phi$ by $J, \tilde{J}$ to match the notation of \cite{CP22}. It is convenient to reorganize the perturbative local operators as follows: for $f \in \C[\![v^{\dot{\alpha}}]\!]$, we denote by $J[f]$ the pairing of the currents $\{J[\vec{m}]\}$, which are valued in $\C[\![v^{\dot{\alpha}}]\!]^*$, with $f$:
\be
	f = \sum\limits_{m_1, m_2 \geq 0} f_{m_1,m_2} (v^{\dot{1}})^{m_1}(v^{\dot{2}})^{m_2} \rightsquigarrow J[f] = \sum\limits_{m_1, m_2 \geq 0} f_{m_1,m_2} J[m_1,m_2]
\ee
and similarly for $\wt{J}$. The generating functions $J[\tilde{\lambda}]$ used in \cite{CP22, CPassoc} correspond to $f = e^{\omega\tilde{\lambda}_{\dot{\alpha}}v^{\dot{\alpha}}}$. With this notation, the non-trivial OPEs of these generators take the following form:
\be
\begin{aligned}
	J_a[f_1](z) J_b[f_2](w) & \sim \frac{f^c{}_{ab}}{z-w} J_c[f_1 f_2](w) \qquad & J_a[f_1](z) \wt{J}^b[f_2](w) & \sim \frac{f^{b}{}_{ca}}{z-w} \wt{J}^c[f_1 f_2](w)\\
\end{aligned}
\ee

The OPEs with $J_a[f]$ are particularly important: they encode the action of infinitesimal, holomorphic (bosonic) gauge transformations.%
\footnote{It is worth noting that the current $\wt{J}$ is \emph{not} a current for gauge transformations. There is no analog of $J$ for the remaining (fermionic) gauge transformations -- the corresponding gauge fields $\BLambda$ were given Neumann bonudary conditions.} %
Recall that the Lie algebra of constant (from the 3d perspective), infinitesimal gauge transformations is the ring of holomorphic maps from $\C^{2}\backslash\{0\}$ to $\fg$, whose bosonic subalgebra is simply $\fg[\![v^{\dot{\alpha}}]\!]$. The group of constant, finite gauge transformations is simply the group $G = G[\![v^{\dot{\alpha}}]\!]$ of series-valued group elements. With this, it is straightforward to verify that the infinitesimal action of gauge transformations is captured by the simple poles in the above OPEs. 

This can be straightforwardly generalized to holomorphic (from the 3d perspective) bosonic gauge transformations: the Lie algebra of infinitesimal, holomorphic gauge transformations can be identified with $\fg_\CO = \fg[\![z; v^{\dot{\alpha}}]\!]$ and the group of finite, holomorphic gauge transformations is the group of invertible series $G_\CO = G[\![z; v^{\dot{\alpha}}]\!]$. As with constant gauge transformations, the above OPEs are invariant under the action of holomorphic gauge transformations.

\subsubsection{Wilson line modules from Koszul duality}
\label{sec:KDmodules}
We can engineer modules for the (celestial) chiral algebras by introducing additional defects transverse to the chiral algebra plane, as suggested in \S 5.3 of \cite{CP22}. In that work it was noted that a holomorphic Wilson defect of the form $\int_{v_1 - \lambda v_2 = -\epsilon} \mathcal{A}$ supported on a noncompact surface of the form $v_1 - \lambda v_2 + \epsilon = 0$ sources a state on twistor space of conformal dimension 1 \footnote{This can be easily verified by acting with the generators of the Virasoro algebra represented as vector fields on twistor space, e.g. $L_0 = -z \partial_z - {1 \over 2} v_i \partial_{v_i}$.} of the form $dz \delta_{z = z_0}{1 \over v_1 + \lambda v_2 + \epsilon}$. (More generally, operators of positive integral conformal dimension are anticipated to be related to Goldstone modes conjugate to the conformally soft tower and have been discussed from a 4d perspective in e.g. \cite{Donnay:2022sdg, Freidel:2022skz}). In this subsection, we will study a closely related family of modules arising from holomorphic Wilson lines. These differ from the Wilson ``defects'' of \cite{CP22} because of the presence of a nontrivial integral kernel, which is essential for constructing a partial connection on a complex manifold like twistor space (see, e.g. \cite{Mason:2010yk, Adamo:2011pv, Bu:2022dis} for more details on the construction of holomorphic Wilson lines and Wilson loops in twistor space).

We will derive the action of the chiral algebra elements on modules corresponding to 6d holomorphic Wilson lines using the technique of Koszul duality; for related analyses of modules by way of Koszul duality\footnote{For the Wilson defects of \cite{CP22}, we need to instead use curved Koszul duality \cite{CPKoszul}.}, see e.g.  \cite{OhZhou, GaiottoOh, GaiottoRapcak}. Here, we will describe the modules realized by Wilson lines from both 3d and 6d realizations of these chiral algebras; in brief, 6d (holomorphic) Wilson lines will reduce to 3d Wilson lines, with the representation of $\mathfrak{g}[v^{\dot{\alpha}}]$ in 3d encoding both the 6d representation and the support of the Wilson line.%
\footnote{We thank A. Sharma for informing us that these modules have also been studied independently by A. Sharma, E. Casali, and W. Bu in work to appear.} %

We will warm up with the 3d Wilson line computation and then generalize to the case of modules arising from a 6d holomorphic Wilson line. Consider a Wilson line intersecting/ending on a universal holomorphic defect in one of the 3d mixed holomorphic-topological gauge theories described in Section \ref{sec:monopoles}. Focusing on the gauge fields, the universal holomorphic defect has local operators $J_a(z)$ and $\wt{J}^a(z)$ coupling to the bulk fields $A, B$ as
\be
	\int J_a \BA^a + \wt{J}^a \BB_a.
\ee
We then consider a local operator $M^R$ on the holomorphic defect that is also attached to a Wilson line $\CW_R$ extended in the topological direction.%
\footnote{We could similarly consider a local operator in the defect plane connecting two Wilson lines, one on either side of the defect plane; we will only consider the case where one of the two Wilson lines is trivial.} %
Requiring this configuration to be gauge invariant then constrains the OPEs of $J_a$ $\wt{J}^a$ with the local operator $M^R$:
\be
\begin{aligned}
	0 & = Q\bigg[\bigg(1+ \int\limits_{z \neq w} \big(J_a(z) \BA^a(z,\overline{z},0) + \wt{J}^a(z) \BB_{a}(z,\overline{z},0)\big) + \dots\bigg)\\
	& \hspace{2cm} \times \bigg(1+ \int\limits_0^\infty \BA^a(w,\overline{w}, t) \rho_a + \dots\bigg)M^R(w) \bigg]\\
	& = \bigg(\int\limits_{z \neq w} \big(J_a(z) (\overline{\partial}\BA^a + \dots) + \wt{J}^a(z) (\overline{\partial}\BB_a + \dots)\big) + \dots\bigg)M^R(w)\\
	& \hspace{2cm} + \bigg(\int\limits_0^\infty (\diff_t \BA^a + \dots) \rho_a + \dots\bigg)M^R(w)\\
	& = \sum_{n \geq 0}\bigg[\bigg(\oint\tfrac{(z-w)^n}{n!} J_a(z) M^R(w) - \delta_{n,0}M^R(w)\bigg) \pd^n_w\BA^a(w)\\
	& \hspace{2cm} + \bigg(\oint\tfrac{(z-w)^n}{n!} \wt{J}^a(z) M^R(w)\bigg) \pd^n_w\BB_a(w)\bigg] + \dots
\end{aligned}
\ee
where $\rho_a$ are the representation matrices for the $\mathfrak{g}$ action on the representation $R$; we used integration-by-parts and Stokes' theorem in the last equality. We see that the OPE of $J_a, \wt{J}^a$ and $M^R$ take the following form:
\be
	J_a(z) M(w) \sim \frac{\rho_a}{z-w} M^R(w) \qquad \wt{J}^a(z) M^R(w) \sim 0.
\ee
Namely, $M^R$ is necessarily a primary operator for the $\mathfrak{g}$ current algebra generated by $J_a$ that has non-singular OPEs with $\wt{J}^a$.

We now turn to the 6d analogue of this computation. There are two important differences. First, unlike the above example, there is a non-trivial choice of support for the 6d Wilson line: along with the location $w$ on the the chiral algebra plane, we must choose a holomorphic curve in the transverse $\C^2$. Additionally, we must include a tower of junction local operators that couple to the fluctuations of the Wilson line in the directions mutually transverse to the chiral algebra plane and the defect. For simplicity, consider placing the Wilson line along the $v^{\dot{1}}$ plane at $v^{\dot{2}} = 0$. The coupling between the junction local operators and the Wilson line is then given by
\be
	\sum\limits_{n \geq 0} \CD_{(0,n)} \CW_R M^R[n] = M^R[0] + \sum\limits_{n} \int \frac{\diff v^{\dot{1}}}{2 \pi i v^{\dot{1}}} \CD_{(0,n)}\CA^a \rho_a M^R[n] + \dots
\ee
where $\CD_{(m_1, m_2)}:= \frac{1}{m_1! m_2!} \pd^{m_1}_{\dot{1}}\pd^{m_2}_{\dot{2}}$. Notice that to define a holomorphic Wilson line we must wedge the gauge field with an appropriate meromorphic differential, which results in a nontrivial integral kernel depending on the support of the Wilson line.

We can derive the module structure by imposing gauge invariance of this coupling together with the universal defect coupling
\be
	\sum\limits_{m_1, m_2} \int J_a[\vec{m}] \CD_{\vec{m}} \CA^a + \wt{J}^a[\vec{m}] \CD_{\vec{m}} \CB_a.
\ee
Due to the similarity to the above computation, we simply state the result: the non-trivial OPEs are given by
\be
	J_a[\vec{m}](z) M^R[n](w) \sim \frac{\delta_{m_1,0} \rho_a}{z-w} M^R[n+m_2](w) \qquad \wt{J}^a[\vec{m}](z) M^R[n](w) \sim 0,
\ee
We see that the tower of operators $M^R[n]$ can be identified with primary operators for the $\mathfrak{g}[v^{\dot{\alpha}}]$ currents $J_a[m_1, m_2]$ induced by the representation $R[v^{\dot{\alpha}}]/(v^{\dot{1}})$. More generally, if we place the Wilson line along the curve $v^{\dot{1}} - \lambda v^{\dot{2}} = 0$ we find the module induced from $R[v^{\dot{\alpha}}]/(\lambda v^{\dot{1}} - v^{\dot{2}})$. We will denote the more general module elements $M^{R,\lambda}[n]$; the parameter $\lambda$ naturally lives on the Riemann sphere $\PP^1$ with the above case corresponding to $\lambda = \infty$. In terms of the notation introduced above, we can write this action as follows:
\be
	J_a[f](z) M^{R,\lambda}[p](w) \sim \frac{\rho_a}{z-w} M^{R,\lambda}[fp](w) \qquad \wt{J}^a[f](z) M^{R,\lambda}[p](w) \sim 0,
\ee
where $p \in \C[\![v^{\dot{\alpha}}]\!]/(\lambda v^{\dot{1}} - v^{\dot{2}})$ and $fp$ denotes the natural action of $f \in \C[\![v^{\dot{\alpha}}]\!]$ on $p \in \C[\![v^{\dot{\alpha}}]\!]/(\lambda v^{\dot{1}} - v^{\dot{2}})$. As mentioned above, we arrive at the conclusion that these 6d holomorphic Wilson lines reduce to Wilson lines in the 3d theory obtained by reduction on $S^3$.

Before moving on, we note the quantum numbers of the operators $M^{R,\lambda}[n]$. For starters, the local operator $M^\lambda[n](w)$ has spin $-\frac{n}{2}$ and scaling dimension $-n$; this follows from the spin and scaling dimension of the $J_a[\vec{m}]$ together with the fact that the Wilson line is rotation invariant. The action of the $SU(2)$ rotating the $v^{\dot{\alpha}}$ as a doublet does not preserve these modules; geometrically, this is because the locus $v^{\dot{1}} - \lambda v^{\dot{2}} = 0$ isn't invariant. Instead, this $SU(2)$ rotates these modules into one another via its natural action on $\PP^1$. The module at $\lambda = \infty$ (resp. $\lambda = 0$) is compatible with the diagonal torus that rotates $J_a[\vec{m}]$, $\wt{J}^a[\vec{m}]$ with weight $\frac{1}{2}(m_1 -m_2)$, with $M^{R,\infty}[n]$ (resp. $M^{R,0}[n]$) given weight $-n$ (resp. $n$). The modules associated to other points of $\PP^1$ also admit a grading coming from the $U(1) \subset SU(2)$ stabilizing $\lambda$, but the operators $J_a[\vec{m}]$, $\wt{J}^a[\vec{m}]$ will no longer be weight vectors.

\subsection{Celestial monopole operators}
\label{sec:QED}

In this subsection, we will describe non-perturbative corrections to the celestial chiral algebra. From the perspective of the above 3d theory, we find such corrections arise from boundary monopole operators. To simplify the discussion, and to avoid the difficulties of non-abelian monopoles described in Section \ref{sec:monopoles}, we will restrict our attention to self-dual abelian gauge theory; to make it somewhat more interesting, we introduce electron and positron fields to realize a self-dual version of QED.

As a holomorphic theory on twistor space, the field content of self-dual QED consists of the gauge fields $\CA \in \Omega^{(0,1)}(\PT)$ and $\CB \in \Omega^{(0,1)}(\PT, \CO(-4)) \simeq \Omega^{(3,1)}(\PT)$ (the components of each are bosonic) as well as the ``electron'' $\CX \in \Omega^{(0,1)}(\PT, \CO(-1)), \CY \in \Omega^{(0,1)}(\PT, \CO(-3))$ and ``positron'' $\ol{\CX} \in \Omega^{(0,1)}(\PT, \CO(-1)), \ol{\CY} \in \Omega^{(0,1)}(\PT, \CO(-3))$. The action for the theory takes the particularly simple form
\be
\label{eq:6dQED}
	S_{\textrm{6d}} = \int \CB \ol{\pd} \CA + \CY \ol{\pd}_\CA \CX + \ol{\CY} ~ \ol{\pd}_\CA \ol{\CX}.
\ee
It is important to note that this theory suffers from the above anomaly and, moreover, this anomaly cannot be cured by a Green-Schwarz mechanism as in the case of the non-abelian gauge theories studied in \cite{CP22, CPassoc}. In particular, we should only expect our non-perturbative boundary chiral algebra to reproduce tree-level results.

\subsubsection{Reduction to 3d and the celestial chiral algebra}
\label{sec:3dQED}
The reduction of the 6d gauge fields $\CA, \CB$ to 3d is the same as in pure self-dual gauge theory. The electron fields produce four towers: the fields arising from $\CY$ will be denoted $\BPsi[m_1,m_2], \BY[m_1,m_2]$ (spins $\frac{1-m_1-m_2}{2}$ and $\frac{3+m_1+m_2}{2}$) and the fields arising from $\CX$ will be denoted $\BPi[m_1,m_2], \BX[m_1,m_2]$ (spins $-\frac{1+m_1+m_2}{2}$ and $\frac{1+m_1+m_2}{2}$). The decomposition of the positron is identical. The fields $\BPsi, \BPi, \ol{\BPsi}, \ol{\BPi}$ are fermionic and the fields $\BX, \BY, \ol{\BX}, \ol{\BY}$ are bosonic. The 6d action in Eq. \eqref{eq:6dQED} reduces to the following action (we suppress the dependence on $v^{\dot{\alpha}}$):
\be
\label{eq:3dQED}
\begin{aligned}
	S_{\textrm{3d}} & = \int \oint\frac{\diff v^{\dot{1}}}{2\pi i}\oint\frac{\diff v^{\dot{2}}}{2\pi i} \bigg(\BB \diff' \BA + \BLambda \diff' \BPhi + \BPsi \diff'_{\BA} \BX + \BPi \diff'_{\BA} \BY + \ol{\BPsi} \diff'_{\BA} \ol{\BX} + \ol{\BPi} \diff'_{\BA} \ol{\BY}\bigg)\\
	& \qquad + \int \oint\frac{\diff v^{\dot{1}}}{2\pi i}\oint\frac{\diff v^{\dot{2}}}{2\pi i} \bigg(\BPhi\big(\BY \BX - \ol{\BY} \ol{\BX}\big)\bigg)
\end{aligned}
\ee

The generalization of the above boundary conditions corresponds to imposing Dirichlet boundary conditions on the photon ``vector multiplets'' $(\BA, \BB)$ and the electron/positron ``chiral multiplets'' $(\BX, \BPsi)$, $(\BY, \BPi)$, $(\ol{\BX}, \ol{\BPsi})$, $(\ol{\BY}, \ol{\BPi})$ while imposing Neumann boundary conditions on photon ``chiral multiplets'' $(\BPhi, \BLambda)$. In analogy with \cite[Section 7.1]{CDG}, the perturbative algebra of local operators is generated by the boundary values of the bosons $B[\vec{m}], \phi[\vec{m}]$ (corresponding to positive/negative helicity gluons) and the fermions $\Psi[\vec{m}]$, $\Pi[\vec{m}]$,$\ol{\Psi}[\vec{m}]$, $\ol{\Pi}[\vec{m}]$ (corresponding to positive/negative helicity electrons and positrons). As before, we denote $B, \phi$ by $J, \tilde{J}$ to match the notation of \cite{CP22}. Using the notation described in the previous section, the non-trivial OPEs of these generators take the following form:
\be
\begin{aligned}
	J[f_1](z) \Psi[f_2](w) & \sim \frac{-1}{z-w} \Psi[f_1 f_2](w) \qquad & J[f_1](z) \ol{\Psi}[f_2](w) & \sim \frac{1}{z-w} \ol{\Psi}[f_1 f_2](w)\\
	J[f_1](z) \Pi[f_2](w) & \sim \frac{1}{z-w} \Pi[f_1 f_2](w) \qquad & J[f_1](z) \ol{\Pi}[f_2](w) & \sim \frac{-1}{z-w} \ol{\Pi}[f_1 f_2](w)\\
	\Pi[f_1](z) \Psi[f_2](w) & \sim \frac{1}{z-w} \tilde{J}[f_1 f_2](w) \qquad & \ol{\Psi}[f_1](z) \ol{\Pi}[f_2](w) & \sim \frac{1}{z-w} \tilde{J}[f_1 f_2](w)
\end{aligned}
\ee

The Lie algebra of constant (from the 3d perspective), infinitesimal (bosonic) gauge transformations is simply $\fg = \C[\![v^{\dot{\alpha}}]\!]$. The group of constant, finite gauge transformations is simply the group of invertible series $G = \C[\![v^{\dot{\alpha}}]\!]^\times$, where the group structure comes from multiplication of series; a series $g(v^{\dot{\alpha}}) \in \C[\![v^{\dot{\alpha}}]\!]$ is invertible if and only if $g(0) \neq 0$. With this, it is straightforward to verify that the action of the constant gauge transformations $g(v^{\dot{\alpha}}) \in G$ is
\be
\begin{aligned}
	g \cdot J[f] & = J[f] \qquad & g \cdot \wt{J}[f] & = \wt{J}[f]\\
	g \cdot \Psi[f] & = \Psi[g^{-1} f] \qquad & g \cdot \ol{\Psi}[f] & = \ol{\Psi}[g f]\\
	g \cdot \Pi[f] & = \Pi[g f] \qquad & g \cdot \ol{\Pi}[f] & = \ol{\Pi}[g^{-1} f]\\
\end{aligned}
\ee
The Lie algebra of holomorphic, infinitesimal gauge transformations can similarly be identified with $\fg_\CO = \C[\![z; v^{\dot{\alpha}}]\!]$ and the group of holomorphic, finite gauge transformations is the group of invertible series $G_\CO = \C[\![z; v^{\dot{\alpha}}]\!]^\times$. As before, a series $g(z;v^{\dot{\alpha}})$ is invertible if it is non-vanishing at $z = v^{\dot{\alpha}} = 0$. The action of this holomorphic gauge transformation follows from the above. For example, the action on $\Pi[f]$ takes the following form:
\be
	g(z;v^{\dot{\alpha}}) = \sum\limits_{n \geq 0} g_n(v^{\dot{\alpha}}) z^n \rightsquigarrow g \cdot \Pi[f] = \sum\limits_{n \geq 0} z^n \Pi[g_n f]
\ee
As with constant gauge transformations, the above OPEs are invariant under the action of holomorphic gauge transformations.

We note that this tree-level chiral algebra has two gradings beyond spin and electric charge $q$: there is the axial charge $q_A$ and the auxiliary $\Z$ grading $a$ of $BF$-like theories that scales the cotangent directions with weight 1 (this is the ``number of $\tilde{J}$s'' grading used in \cite[Lemma 9.0.1]{CP22}). Note that the latter grading does not survive the deformation away from self-dual gauge theory.

\begin{table}[h!]
	\centering
	\begin{tabular}{c|c|c|c|c|c|c}
		& $J$   & $\tilde{J}$  & $\Psi$ & $\Pi$ & $\ol{\Psi}$ & $\ol{\Pi}$ \\ \hline
		$q$ & $0$ & $0$ & $-1$ & $1$ & $1$ & $-1$\\
		$q_A$ & $0$ & $0$ & $-1$ & $1$ & $-1$ & $1$\\
		$a$ & $0$ & $1$ & $1$ & $0$ & $1$ & $0$\\
	\end{tabular}
	\caption{Electric charge ($q$), axial charge ($q_A$), and auxiliary grading ($a$) of the generators of perturbative celestial chiral algebra for tree-level self-dual QED.}
	\label{table:}
\end{table}

\subsubsection{Celestial spectral flow in self-dual QED}
We now move to non-perturbative operators in the celestial chiral algebra. From the perspective of this chiral algebra as living on the boundary of a 3d theory, these non-perturbative corrections will correspond to boundary monopole operators.

As mentioned above, the group of holomorphic gauge transformations in our 3d incarnation of self-dual Maxwell theory is the group of invertible series in the three coordinates $z, v^{\dot{\alpha}}$. From the perspective of the celestial chiral algebra, these holomorphic gauge transformations preserve the vacuum module. That said, the algebra of modes is invariant under a larger symmetry group: the group of meromorphic (from the 3d perspective) gauge transformations $G_\CK = \C(\!(z)\!)[\![v^{\dot{\alpha}}]\!]^\times$; such a Laurent series $g(z;v^{\dot{\alpha}})$ is invertible so long as the coefficient of the lowest power of $z$ is an invertible series in $v^{\dot{\alpha}}$.

The quotient of the group of meromorphic gauge transformations by the group of holomorphic gauge transformations is the affine Grassmannian $\textrm{Gr}_G = G_\CK/G_\CO$. It is fairly straightforward to see that the (closed points of the) affine Grassmannian for $G = \C[\![v^{\dot{\alpha}}]\!]^\times$ can be identified with $\Z$: let $g(z;v^{\dot{\alpha}}) \in G_\CK$ be any meromorphic gauge transformation and let $\m_0$ by the smallest power of $z$ with a non-vanishing (and necessarily invertible) coefficient. It follows that $g(z;v^{\dot{\alpha}})$ and $z^{\m_0}$ differ by the holomorphic gauge transformation $g' = z^{-\m_0} g$ and, moreover, there are no non-trivial elements in $G_\CO$ that preserve $z^{\m_0}$, whence the quotient is $\Z$.

In analogy with the discussion in Section \ref{sec:bdymonopoles}, the point $z^\m$ in the affine Grassmannian $\textrm{Gr}_G$ gives a spectral flow operation $\sigma_\m$. The action of this spectral flow is realized by applying the meromorphic gauge transformation $z^\m$:
\be
\label{eq:SDQEDopes}
\begin{aligned}
	\sigma_\m(J[f]) & = J[f] \qquad & \sigma_\m(\wt{J}[f]) & = \wt{J}[f]\\
	\sigma_\m(\Psi[f]) & = z^{-\m} \Psi[f] \qquad & \sigma_\m(\ol{\Psi}[f]) & = z^\m \ol{\Psi}[f]\\
	\sigma_\m(\Pi[f]) &= z^\m \Pi[f] \qquad & \sigma_\m(\ol{\Pi}[f]) &= z^{-\m} \ol{\Pi}[f]
\end{aligned}
\ee
We will denote the local operator arising from spectral flow of the vacuum vector by $V_\m$; the integer $\m$ encodes the 3d magnetic charge/monopole number of the monopole. We take as our ansatz that the full, non-perturbative chiral algebra is realized by extending the perturbative celestial chiral algebra by these modules. Once we extend the celestial chiral algebra by these generators, we gain another $\Z$ grading by 3d magnetic charge $\m$; the perturbative chiral algebra corrresponds to the subalgebra of charge $0$.

We expect these boundary monopoles to serve two roles in the same way as outlined in Section \ref{sec:abelian3dbdy}. First, they should enforce the quantization of electric charge: modules for the full, non-perturbative celestial chiral algebra must have integral electric charges, whereas the perturbative analysis described in Section \ref{sec:KDmodules} allows Wilson lines with arbitrary complex charge. At the level of OPEs, we find
\be
	V_\m(z) M^{q,\lambda}[p](w) \sim (z-w)^{q\m} \sigma_\m(M^{q,\lambda}[p](w)) + \dots
\ee
whence $q \m$ must be integral. Second, these celestial monopoles will identify modules for the perturbative algebra that differ from one another by spectral flow. We note that because the abelian currents in the celestial chiral algebra have vanishing level the boundary monopoles have vanishing electric charge and therefore do not identify modules with different electric charges as in 3d Chern-Simons theories.

\subsubsection{4d interpretation}
\label{sec:4dinterpret}
We attempt to understand the interpretation of the 4d states corresponding to these spectral flow modules by way of the 6d bridge established in \cite{CP22}. We start by noting that the states in the $\m$th spectral flow of the vacuum are in 1-to-1 correspondence with the states in the vacuum module, but their spins are shifted: for example, we saw $\sigma_\m(\Psi[f]) = z^{-\m} \Psi[f]$ and so the spin of this field, and hence its modes $\Psi[m_1, m_2]_n$ and the states they generate, are shifted by $\m$ -- the field $\Psi[m_1, m_2]$ has spin $\frac{1-m_1-m_2}{2}$ whereas its $\m$th spectral flow has spin $\m + \frac{1-m_1-m_2}{2}$. The quantum numbers $m_1, m_2$ are unchanged -- our spectral flow morphisms only alter the quantum numbers for the half of the 4d spin group $\textrm{Spin}(4) \simeq SU(2) \times SU(2)$ rotating the twistor sphere.

Importantly, we see that the resulting states are precisely those arising from reducing a 6d field $\sigma_\m(\CY)$ valued in $\CO(-2\m-3)$, corresponding to a massless 4d field of helicity $-\tfrac{1}{2}-\m$. More generally, spectral flow will take a 6d field $\CZ$ valued in $\CO(2h-2)$ with electric charge $q$ to another 6d field $\sigma_\m(\CZ)$ valued in $\CO(2h-2+2q\m)$. Thus, the states in the $\m$th spectral flow module can be identified with states resulting from perturbing around the holomorphic gauge bundle $\CO(2\m) \to \PT$ in 6d.%
\footnote{Rather, we perturb around the principal $\C^\times$ bundle $P_{2\m}$ for which $\CO(2\m)$ is the line bundle associated to the charge 1 representation $\CO(2\m) = P_{2\m} \times_{\C^\times} \C$.} %
Coupling to this bundle results in an apparent shift in helicity $h \to h+q\m$. 

The shift in helicity due to the presence of a non-trivial gauge bundle is a familiar phenomenon: this is exactly the shift in angular momentum experienced by a electrically charged particle in the presence of a magnetic monopole \cite{Zwanziger:1972sx, Csaki:2020inw, Csaki:2022tvb}! Somewhat more precisely, one should say that a 2-particle state of an electrically charged particle and a magnetically charged particle has a non-trivial ``pairwise helicity.'' We understand this as follows, see, e.g., \cite[Section 2]{CHSTTW20} for a detailed discussion of these notions.%
\footnote{This reference uses a convention where the pairing of electric and magnetic charges is half-integral, i.e. our charge $\m$ is twice their charge $g$.} %
If we boost to the center of momentum frame, or any frame where the momenta of the particles point towards antipodal points on the celestial sphere, it is easy to see that any such configuration is preserved by an $SO(2)$ little group rotating a transverse plane; 2-particle states are thus labeled by a pair of 1-particle data as well as an additional quantum number describing its possible transformations under this ``pairwise little group.'' Namely, under such a rotation by angle $\phi$, a general 2-particle state acquires a phase $e^{i(s_1 + s_2 + h_{12})\phi}$, where $s_i$ are the spins/helicities of the two particles and $h_{12}$ is the aforementioned pairwise helicity. The case of $h_{12} = 0$ describes the direct product of two 1-particle states. More generally, if particle 1 has electric charge $q \in \Lambda_{\rm weight}$ and particle 2 has magnetic charge $\m \in \Lambda_{\rm weight}^\vee$, then one finds exactly $h_{12} = \tfrac{1}{2}\langle \m, q \rangle$.

In the present setting, we interpret the insertion point of the operator $V_\m$ as a point on the celestial sphere; placing a another operator of electric charge $q$ at $z$, we see that the shift in spin by $q\m$ precisely accounts for the expected additional phase accrued when rotating the celestial sphere. Curiously, the natural spectral flows $\sigma_\m$ appear to realize even magnetic charges, corresponding to the fact that they realize a coupling to $\CO(2\m)$. There are spectral flow morphisms for the remaining magnetic charges, corresponding to coupling to the bundles $\CO(2\m+1)$, but these lead to twisted modules, cf. the spectral flow maps exchanging Ramond and Neveu-Schwarz sectors in 2d superconformal theories, and hence cannot be used to extend the celestial chiral algebra. It would be interesting to see if these twisted modules can be related to the twisted sectors proposed by \cite{vanBeest:2023dbu} to describe out-states in the scattering of charged particles off of heavy monopoles.

Absent this shift in helicity, the 4d interpretation of our celestial chiral algebra generators $V_\m$ is not immediately clear. An interesting and immediate question raised by our 3d magnetically charged operators realized by spectral flow is whether there are magnetic analogs $J^\vee$ of the currents $J$ that implement Strominger's magnetic soft theorem \cite{Strominger:2015bla} (or the non-abelian extension proposed in \cite{Kapec:2021eug}). In the case of pure abelian self-dual gauge theory, it is clear that there can be no such operator without adding additional generators to the chiral algebra: the operators $V_\m(z)$ corresponding to the states $\sigma_\m(\ket{0})$ have regular OPEs with everything. A tantalizing possibility is that there is such an operator after the inclusion of suitable Goldstone fields that are canonically conjugate to the $J$'s, cf. Section 3 of \cite{Donnay:2018neh}. As seen in \cite{Crawley:2023brz}, suitable coherent states of these Goldstone fields form interesting modules in the gravitational celestial chiral algebra realizing self-dual Kerr Taub-NUT backgrounds. In the case of pure self-dual abelian gauge theory, there are analogous Goldstone fields $S$ that are canonically conjugate to the $J$s; the modules realized as coherent states of these fields lead to Faddeev-Kulish dressings factors \cite{Kulish:1970ut, Chung:1965zza}, cf. Section 3.3 of \cite{Arkani-Hamed:2020gyp}; see also \cite{Choi:2018oel}. Our monopoles $V_\m$ serve a complementary role where the monopole modules are coherent states built from $J$, cf. 3d monopoles are vertex operators $e^{\m \varphi}$ for the current $J = \pd \varphi$, rather than from $S$. As the Goldstone modes $S$ are canonically conjugate to the $J$'s, there is an OPE
\be
	S[0,0](z) J[0,0](w) \sim \frac{1}{(z-w)^2}
\ee
and hence
\be
	S[0,0](z) V_\m(w) \sim \frac{m V_\m(w)}{z-w}
\ee
We see that the $V_\m(w)$ are charged under the symmetry generated by Goldstone boson $S[0,0]$. Such Goldstone modes \footnote{An infinite tower of such modes of positive integer conformal weight, symplectically paired with the conformally soft modes comprising the perturbative chiral algebra, has been explored recently in \cite{Freidel:2022skz}.  There it was shown that suitably constructed Goldstone-dressed states trivialize the full tower of conformally soft theorems. It will be interesting if these properties can help us better understand the 4d interpretation of the spectral flow of the 2d vacuum.} in gauge theory generate a large gauge transformation at null infinity, resulting in a shift of the 4d vacuum. The nontrivial action on the spectral flow vertex operator is natural, as the $V_{\m}$ should interact with the nontrivial 4d background of QED created by the Goldstone.

Although these operators carry 3d magnetic charge, they are not the right candidates for constructing the dual ``magnetic'' soft currents in 4d. As pointed out by Footnote 16 of \cite{Arkani-Hamed:2020gyp}, further incorporating the magnetic soft theorem seems to require a further extension or doubling of the algebra generated by $J, S$; how this could arise in a twistorial set-up is an open question.

Another perspective on our monopoles $V_\m$ comes from reducing the 6d gauge bundles down to 4d. If we start with the bundle $\CO(\m) \to \PT$, we can attempt to pass it to 4d through the correspondence between $\PT$ and (complexified) spacetime. We consider the correspondence space $\mathbb{F} = \PP^1 \times \R^4$ (or $\PP^1 \times \C^4$); this space fits into a natural diagram:
\be
	\PT \overset{\pi_1}{\longleftarrow} \mathbb{F} \overset{\pi_2}{\longrightarrow} \R^4
\ee
where $\pi_1([z_\alpha], x^{\dot{\alpha} \alpha}) = [x^{\dot{\alpha}\alpha} z_\alpha, z_\alpha]$ and $\pi_2([z_\alpha], x^{\dot{\alpha} \alpha}) = (x^{\dot{\alpha}}).$ It is straightforward to check that pulling back $\CO(\m) \to \PT$ to the correspondence space $\mathbb{F}$ simply results in the product of the trivial bundle on $\R^4$ and the bundle $\CO(\m) \to \PP^1$. Pushing forward this bundle along $\pi_2$ is somewhat more delicate: we to compute the pushforward along $\pi: \PP^1 \to \textrm{point}$ of the bundle $\CO(m)$, which naively computes the space of global sections. Of course, because $\PP^1$ isn't affine, we should really consider the \emph{derived} pushforward, which computes sheaf cohomology. We conclude that the pushforward of $\pi_2^* \CO(\m)$ along $\pi_1$ produces $H^\bullet(\PP^1, \CO(m))$ copies of the trivial bundle on $\R^4$. A straightforward computation shows that this cohomology is
\be
	H^i(\PP^1, \CO(\m)) = \begin{cases}
		\C^{\m+1} & i = 0, \m \geq 0\\
		\C^{|\m+1|} & i = 1, \m < 0\\
		0 & \textrm{else}
	\end{cases}
\ee

Translating this computation to the twistor correspondence, the 6d bundle $\CO(\m)$ on $\PT$ becomes (a suitable symmetric power of) the positive chirality spin bundle $S_+$ or its dual $S_+^*$:%
\footnote{We thank L. Mason for explaining the identification with the spinor bundle to us.} %
\be
	\pi_1{}_* (\pi_2^*\CO(\m)) = \begin{cases}
		\textrm{Sym}^\m S_+ & \m \geq 0\\
		\textrm{Sym}^{-\m-1} S_+^* & \m \leq -2\\
		\underline{0} & \m = -1\\
	\end{cases}
\ee
The easiest way to see this is to note that sheaf cohomology of of $\CO(\m)$ for non-negative $\m$ is identified with degree $\m$ polynomials in $z_\alpha$; Serre duality says $\CO(\m)$ for $\m \leq -2$ is the dual of $\CO(-\m-1)$; and the bundle $\CO(-1)$ has no cohomology and hence reduces to a rank $0$ vector bundle, i.e. the constant sheaf $\underline{0}$ whose fiber is the $0$-dimensional vector space $0$. 

Our results have so far been at the level of associated bundles to the principle $U(1)$ bundle, and we would like to understand things directly in terms of the principle bundle. A major difficulty in relating this bundle to a familiar 4d gauge field configuration per the usual twistor correspondence is the fact that this gauge bundle has nonvanishing Chern class over the twistor sphere \textit{at each point} in spacetime. Sparling \cite{sparling1977dynamically} has explored extensions of the twistor correspondence in the non-abelian case, and where the Chern class is nonvanishing at isolated points in spacetime, which he calls jumping points. There, he finds singularities in the gauge field, which lead to a breaking of the Yang-Mills symmetry.\footnote{In our case, the configuration may therefore be singular at every point in spacetime, and it is not clear to us whether such a solution exists or is stable.} In such cases, Sparling shows that one can obtain solutions to the Yang-Mills equations for symmetry subgroups of the original gauge group, plus certain on-shell massless matter fields. For instance, simple jumping loci for $SU(2)$ result in solutions to two self-dual Maxwell equations plus additional charged fields under the two gauge groups. It is unclear to us how to generalize his reasoning to our situation, but we might hope that our principal $U(1)$ bundle emerges by a similar ``symmetry breaking'' phenomenon, and perhaps could be obtained by reducing the structure group of the spin bundle to the diagonal $U(1) \subset SU(2)_+$.

\subsubsection{Self-dual monopoles}
\label{sec:selfdualmonopole}
Before moving on, it important to mention that there have already been investigations into self-dual field configurations with 4d magnetic charge. Magnetic monopoles in self-dual abelian gauge theory cannot solely have magnetic charge: the self-duality equation implies that there must be an accompanying electric charge so that $E + i B = 0$. Thus, it is reasonable to ask for a holomorphic line bundle on (an open set of) twistor space $\PT$ that realizes such a 4d ``self-dual'' monopole. Thankfully, such a line bundle is known by work of Penrose and Sparling \cite{PS79}; see also Chapter 8.4 of \cite{WWtwistor} and references therein.%
\footnote{We thank A. Sharma for explaining aspects of the so-called ``twistor quadrille'' to us.} %

We start with a self-dual monopole with worldline along the $x^4$ axis. The corresponding field strength can be realized by taking the twistor function $\ln(v^{\dot{2}} - \frac{v^{\dot{1}}}{z})$; passing this twistor function through the Penrose transform yields a field strength with self-dual part
\be
	F_{\dot{\alpha}\dot{\beta}} = \frac{1}{4}\begin{pmatrix}
		\frac{x^1+i x^2}{r^3} & \frac{x^3}{r^3}\\
		\frac{x^3}{r^3} & \frac{x^1-i x^2}{r^3}
	\end{pmatrix}
\ee
where $r^2 = (x^1)^2 + (x^2)^2 + (x^3)^2$, as desired. This twistor function admits a natural generalization to arbitrary momentum $p_{\alpha \dot{\alpha}}$ by replacing the twistor function with
\be
	f_p(z,v^{\dot{\alpha}}) = \ln\bigg(\bigg(p_{2 \dot{\alpha}} - \frac{p_{1 \dot{\alpha}}}{z}\bigg) v^{\dot{\alpha}}\bigg)
\ee
The corresponding self-dual field strength then takes the form
\be
	F_{\dot{\alpha}\dot{\beta}} = \frac{i p^2}{4}\begin{pmatrix}
		\frac{(p_1 + i p_2)(x^3 + i x^4) + (p_3 + i p_4)(x^1 + i x^2)}{\big(p^2 x^2 - (p\cdot x)^2\big)^{3/2}} & \frac{i\big(p_1 x^2 - p_2 x^1 + p_3 x^4 - p_4 x^3\big)}{\big(p^2 x^2 - (p\cdot x)^2\big)^{3/2}}\\
		\frac{i\big(p_1 x^2 - p_2 x^1 + p_3 x^4 - p_4 x^3\big)}{\big(p^2 x^2 - (p\cdot x)^2\big)^{3/2}} & \frac{(p_1-i p_2)(x^3 - i x^4) + (p_3 - i p_4)(x^1 - i x^2)}{\big(p^2 x^2 - (p\cdot x)^2\big)^{3/2}}
	\end{pmatrix}
\ee
This field strength is invariant under scaling $p \to \lambda p$ for $\lambda \in \R_{>0}$, so it only depends on the worldline and not the momentum, and vanishes if we choose a null momentum $p^2 = 0$.

We would like to ask how the insertion of one of these self-dual monopoles alters the celestial chiral algebra. The twistor function $f_p$ above should be interpreted as the logarithm of a single transition function describing an underlying line bundle $\CL_p$ on $\PT$; although we will not describe them, the line bundles $\CL_p$ are constructed explicitly in \cite{PS79}. The construction of the bundle in \cite{PS79} works twistor sphere by twistor sphere: for each twistor sphere away from the worldline, they introduce a four-set open cover and glue them with certain holomorphic transition functions. 

Unfortunately, we do not expect the line bundle $\CL_p$ admits an algebro-geometric description. Indeed, it is a (complex) line bundle over a non-complex submanifold of $\PT$: it is defined on the complement of $\mathbb{R} \times \mathbb{P}^1$ in $\mathbb{PT}$. Although this manifold is not complex, it inherits a structure from the holomorphic structure on twistor space. The transition functions of the quadrille are compatible with this inherited structure. Because of the compatibility between the transition functions for $\CL_p$ and the inherited complex structure, it is possible to define Dolbeault cohomology with values in $\CL_p$.

Reducing this 6d field configuration to 3d would involve computing tangential Cauchy-Riemann cohomology of $S^3\backslash\{N,S\} \subset \PT\backslash (\R \times \PP^1)$ with values in $\CL_p$, cf. \cite{Zeng2023}. We expect this should introduce some sort of module for the celestial chiral algebra whose structure depends on the precise insertion point of the module on the chiral algebra plane, thereby encoding the 4d support of the self-dual monopole. We expect that this module, arising from the insertion of a charged source in spacetime, will have similar features to the configuration in self-dual gravity studied in \cite{Guevara:2021yud}; in both gauge theory and gravity, it would be desirable to have a description of the module that allows us to write down the action of all modes of the perturbative chiral algebra.

One structural aspect of this module is that it should not be preserved by either the $SU(2)$ rotating the twistor sphere or the $SU(2)$ rotating the KK modes. Indeed, the worldline of the self-dual monopole isn't invariant under all of $Spin(4) \simeq SU(2) \times SU(2)$: it is only invariant under a $Spin(3) \simeq SU(2)$ subgroup thereof that fixes the worldline (i.e. the subgroup rotating the linking sphere). Correspondingly, this should be a feature for \emph{any} module associated with a 4d line operator. In the example where the worldline is along the $x^4$ axis, we should only expect to preserve the diagonal $SU(2)$ subgroup. As a result, elements in such a module would need a compensating translation in the fiber directions $v^{\dot{\alpha}}$ to be invariant under translations in the chiral algebra $z$ plane: the worldline is only invariant under $\pd_z + v^{\dot{2}}\pd_{\dot{1}}$, corresponding to the translation $(z,v^{\dot{1}}, v^{\dot{2}}) \to (z-w, v^{\dot{1}}-w v^{\dot{2}}, v^{\dot{2}})$. Similarly, the action of rotations is modified so that $v^{\dot{1}}$ has spin $1$ and $v^{\dot{2}}$ has spin 0; if we expand an element of this module into modes $M[\vec{m}]$, the difference in spin between $M[\vec{m}]$ and $M[\vec{0}]$ is $-m_1$, rather than the usual $-\frac{m_1+m_2}{2}$.

\section{A First Step Towards Non-abelian Monopoles}\label{sec:nonabelian}
We saw in the previous section that there are interesting non-perturbative corrections to the celestial chiral algebra corresponding to unusual 4d helicity-shifting field configurations or 3d boundary monopoles. Of course, the discussion in the previous section was restricted to tree-level, self-dual abelian gauge theories. This restriction was entirely technical: we do not currently have the tools to perform the full, non-perturbative analysis of boundary chiral algebras on Dirichlet boundary conditions in 3d non-abelian gauge theories, let alone when the gauge group is an infinite-dimensional beast like $G[\![v^{\dot{\alpha}}]\!]$. That said, if we are interested in going beyond tree-level computations, we must consider these non-abelian gauge theories. 

In this section, we provide a first pass at the non-perturbative corrections present in the celestial chiral algebras of non-abelian theories, focusing on the case of $\fg = \fsl(2)$, with the hope that a more thorough analysis can be performed once the necessary 3d tools are developed. We introduce a family of spectral flow automorphisms that, upon lifting to $\PT$ and passing through the Penrose-Ward correspondence, realize 4d field configurations with non-trivial 4d magnetic charge. These automorphisms are labeled by a 4-vector $p$ identified with the momentum of a 4d monopole, and we provide several consistency checks of this identification.

\subsection{Monopoles from holomorphic bundles on $\PT$}
\label{sec:monopolebundles}
We now briefly review the construction of 4d monopoles from holomorphic bundles on twistor space, in the simplest example. We will focus on the case of $G = SU(2)$.

Consider a 4d monopole with magnetic charge 1 with worldline along the $x^4$ axis; the corresponding holomorphic bundle on $\PT$ is described in Example 8.4.1 of \cite{WWtwistor} and we closely follow the discussion in Chapter 8 of loc. cit. The holomorphic transition function is given by
\be
	g_4(z, v_{\dot{\alpha}}) = \begin{pmatrix}
		z e^{\gamma_4} & \gamma_4{}^{-1}(e^{\gamma_4} - e^{-\gamma_4})\\
		0 & z^{-1} e^{-\gamma_4}
	\end{pmatrix}\,,
\ee
where $\gamma_4 = \frac{i}{2}(v^{\dot{2}} - v^{\dot{1}}/z)$. We now review how to extract a gauge field from this transition function.

First, we restrict the above gauge transformation to the $\PP^1$ in twistor space above the point $x^\mu$ by imposing the incidence relation. If this is to define a non-singular gauge field at $x^\mu$, the corresponding bundle over $\PP^1$ must be trivial, i.e. there must be gauge transformations $\widehat{g}$ (holomorphic away from $z = \infty$) and $\widetilde{g}$ (holomorphic away from $z = 0$) so that $g_4 = \widetilde{g} \widehat{g}^{-1}$ (at $x^\mu$).

The splitting matrices for a wide class of examples, including the present one, are derived in Section 8.2 of \cite{WWtwistor} but we do not need the precise form. Instead, we describe the conditions for the existence of such a splitting. Note that upon restriction to the $\PP^1$ above $x^\mu$, the exponent $\gamma_4$ becomes
\be
	\gamma_4 = x^3 - \tfrac{i}{2}(z+z^{-1})x^2 - \tfrac{1}{2}(z - z^{-1}) x^1\,,
\ee
which is independent of $x^4$. We start by choosing a splitting $\gamma_4 = \widetilde{h}_4 - \widehat{h}_4$ into a part $\widehat{h}_4$ holomorphic away from $z = \infty$ and a part $\widetilde{h}_4$ holomorphic away from $z = 0$, say
\be
	\widetilde{h}_4 = \tfrac{1}{2} x^3 +\tfrac{1}{2}(x^1 - i x^2) z^{-1} \,, \qquad \widehat{h}_4 = -\tfrac{1}{2}x^3 + \tfrac{1}{2}(x^1+ix^2)z\,.
	\ee
	The existence of the desired splitting matrices then depends on the coefficient of $z^0$ of the Laurent series expansion of
	\be
	e^{-\widetilde{h}_4 - \widehat{h}_4} \frac{e^{\widetilde{h}_4 - \widehat{h}_4} - e^{\widehat{h}_4 - \widetilde{h}_4}}{\widetilde{h}_4 - \widehat{h}_4} = \frac{e^{-2\widehat{h}_4} - e^{-2\widetilde{h}_4}}{\widetilde{h}_4 - \widehat{h}_4} = -\sum_{n \geq 1}\frac{(-2)^n}{n!} \sum_{m=0}^{n-1} \widehat{h}_4{}^m \widetilde{h}_4{}^{n-m-1}\,.
\ee
In particular, the points $x^\mu$ where this series coefficient vanishes are exactly where the splitting matrices do not exist or, equivalently, where the 4d gauge field is singular.

It is straightforward to extract this series coefficient: it is given by
\be
	F_4 = - \sum_{\ell \geq 0}\frac{(-2)^{2\ell+1}}{(2\ell+1)!} \big(\tfrac{r}{2}\big)^{2\ell} = \frac{2\sinh(r)}{r}\,,
\ee
where $r^2 = (x^1)^2+(x^2)^2+(x^3)^2$. Note that this never vanishes for real $x^1, x^2, x^3$ and so this holomorphic bundle on $\PT$ induces \emph{non-singular} gauge fields in 4d. In terms of this series coefficient, the 4d gauge field can be expressed in \emph{$R$-gauge} as follows:
\be
	A_{\dot{1} \alpha} = \frac{1}{2 F_4} \begin{pmatrix}
		\CD_{\dot{1} \alpha}F_4 & 0\\ 2 \CD_{\dot{2} \alpha} F_4 & -\CD_{\dot{1} \alpha}F_4\\
	\end{pmatrix} \qquad A_{\dot{2} \alpha} = \frac{1}{2 F_4} \begin{pmatrix}
		-\CD_{\dot{2} \alpha}F_4 & 2 \CD_{\dot{1} \alpha} F_4\\ 0 & \CD_{\dot{2} \alpha}F_4\\
	\end{pmatrix} 
\ee
In these expressions $\CD_{\alpha \dot{\alpha}}$ denotes the covariant derivative for the holomorphic line bundle with transition function $e^{\gamma_4}$, i.e. $\CD_\mu = \partial_\mu - i \delta_{\mu 4}$.

Note that, because $\gamma_4$ and $F_4$ are independent of $x^4$, it follows that the corresponding 4d gauge field $A_{\dot{\alpha} \alpha}$ is independent of $x^4$. Plugging this into the the self-duality equations, it follows that
\be
	B_a = \tfrac{1}{2}\epsilon_{abc}F^{bc} = D_a A_4,
\ee
where $D_a = \pd_a - A_a$ is the covariant derivative for $A_a$, $a = 1,2,3$. These are the equations for a BPS monopole (with $A_4$ identified with the Higgs field)! Explicitly, we find
\be
	A_a = \tfrac{i}{2 r^2}(r \coth r - 1) \epsilon_{abc} x^a \sigma^b + \tfrac{i}{2} \sigma_a\,, \qquad A_4 = \tfrac{i}{2r^2}(r \coth r - 1) x^a \sigma_a\,.
\ee

\subsection{Celestial spectral flow}
\label{sec:spectralflow}
We saw in the previous section how to encode a monopole with worldline given by the $x^4$ axis into a certain holomorphic bundle on twistor space with transition function $g_4$. The states in the presence of the above holomorphic gauge bundle can be obtained by simply applying the above gauge transformation $g_4$.

For example, consider the state corresponding to the field configuration
\be
	\BA(v^{\dot{\alpha}}) = \delta(z-z_0) f(v^{\dot{\alpha}}) t_a \quad \longleftrightarrow \quad \BA[m_1,m_2] = \delta(z-z_0) f_{m_1,m_2} t_a
\ee
for $t_a$, $a = 1,2,3$, a basis of $\mathfrak{sl}(2)$; gauge transformations act on this state by conjugating this field configuration. It will be convenient to consider the Chevalley basis $t_a = \{E,F,H\}$. The state with $t_a = E$ has the easiest transformation:
\be
	\delta(z-z_0) f E \to \delta(z-z_0) \bigg((z_0)^2 e^{2\gamma_4} f\bigg) E\,.
\ee
The next easiest is that of $t_a = H$:
\be
	\delta(z-z_0) f H \to \delta(z-z_0)\bigg[f H + \bigg(2 z_0 f \frac{(1 - e^{2\gamma_4})}{\gamma_4}\bigg) E\bigg]
\ee
Finally, the most involved transformation is that of $t_a = F$:
\be
\begin{aligned}
	\delta(z-z_0) f F & \to  \delta(z-z_0)\bigg[ \big((z_0)^{-2} e^{-2\gamma_4}f\big)F + \bigg((z_0)^{-1} f \frac{(1 - e^{-2\gamma_4})}{\gamma_4}\bigg)H\\
	& \hspace{2.5 cm} + \bigg(f \frac{(2 - e^{2\gamma_4} - e^{-2\gamma_4})}{\gamma_4{}^2}\bigg)E\bigg]
\end{aligned}
\ee

We now translate this transformation of states into a statement about the transformation properties of the (perturbative) generators of the celestial chiral algebra associated to the $SU(2)$ holomorphic BF theory on twistor space (equivalently, self-dual Yang-Mills theory in 4d).%
\footnote{We work at tree-level in this section; see \cite{CP22, CPassoc} for details on the failure of associativity in this chiral algebra at loop-level, and how to remove this failure by cancelling the 1-loop twistor space gauge anomaly \cite{Costello:2021bah}} %
Because the propagator connects $\BA^a[m_1,m_2]$ and $\BB_a[m_2,m_1]$, the above field configuration is sourced by a boundary insertion of
\be
	\BB_a[f] = \sum f_{m_1,m_2} \BB_a[m_1,m_2]
\ee
As above, we denote the 0-form component of this local operator by $J_a[f]$. (The higher form components are never $Q$-closed and trivialize the $t$ and $\bar{z}$ dependence of this operator in cohomology.) The remaining 1-particle states of interest correspond to boundary insertions of
\be
	\BPhi^a[f] := \sum f_{m_1,m_2} \BPhi^a[m_1,m_2]
\ee
which sources the fields $\BLambda_a[m_1,m_2]$. The 0-form component will be denoted by $\tilde{J}_a[f]$ to match the above. %
Translating the above action of large gauge transformations on states to the generators $J, \tilde{J}$ results in a spectral flow automorphism $\sigma_4$ of the perturbative chiral algebra. For brevity, we only write down the action on $J$; the action on $\tilde{J}$ can be obtained by replacing $J \to \tilde{J}$.

First, the generator $J_E[\tilde{\lambda}]$ is mapped as
\be
	J_E[f] \to \sigma_4(J_E)[f] = z_0{}^2 J_E[e^{2\gamma_4} f]\,,
\ee
where we have suppressed the dependence of the current $J$ and $\gamma_4$ on the insertion point $z_0$. Similarly, the spectral flow of $J_H[f]$ is given by
\be
	\sigma_4(J_H)[f] = J_H[f] + 2 z_0 J_E[\gamma_4{}^{-1}(1-e^{2\gamma_4})f]\,.
\ee
For $J_F[f]$ we find
\be
\begin{aligned}
	\sigma_4(J_F)[f] & = z_0{}^{-2} J_F[e^{-2 \gamma_4} f] + z_0^{-1} J_H[\gamma_4^{-1}(1-e^{-2\gamma_4})f] + J_E[\gamma_4{}^{-2}(2-e^{2\gamma_4}-e^{-2\gamma_4})f]\,.
\end{aligned}
\ee

Although we have presented a relatively compact expression for the above spectral flow morphism, it is important to note that it has a rather non-trivial expression in terms of the component fields $J_a[m_1,m_2]$. Even the simple action on $J_E[f]$ involves an infinite number of fields when expressed in term of the component fields (conformally soft modes):
\be
	\sigma_4(J_E[m_1,m_2]) = (z_0)^2 \sum_{n_1,n_2 \geq 0} \frac{(-i/z_0)^{n_1} (i)^{n_2}}{n_1! \, n_2!} J_E[m_1+n_1,m_2+n_2]
\ee
We note that if we set to zero $J_a[m_1,m_2]$ with $m_1,m_2 \neq 0$ (this is equivalent to restricting to $v^{\dot{\alpha}} = 0$) we see that the spectral flow morphism takes the following form:
\be
	v^{\dot{\alpha}} = 0 \rightsquigarrow \begin{cases} \sigma_4(J_E) = z_0{}^2 J_E\\ \sigma_4(J_H) = J_H - 4 z_0 J_E\\ \sigma_4(J_F) = z_0{}^{-2} J_F +2 z_0{}^{-1} J_H - 4 J_E\\ \end{cases}
\ee
This is the usual spectral flow automorphism in Section \ref{sec:bdymonopoles} combined with a holomorphic gauge transformation:
\be
	g_4(z;0) = \begin{pmatrix}
		z & -2\\ 0 & z^{-1}
	\end{pmatrix} = \begin{pmatrix}
		1 & -2z\\ 0 & 1
	\end{pmatrix} \begin{pmatrix}
		z & 0\\ 0 & z^{-1}
	\end{pmatrix}
\ee

\subsection{Scaling and rotations of the twistor sphere}
Note that the spectral flow $\sigma_4$ has neither a definite scaling dimension nor a definite spin, but this should be no surprise -- the 4-momentum $p = (0,0,0,1)$ parameterizing the chosen holomorphic bundle is not preserved by scaling $\R^4$ or rotating the twistor $\PP^1$. More generally, we find bundles parameterized by a general momentum
\be
	p_{\alpha \dot{\alpha}} = p_\mu (\sigma^\mu)_{\alpha \dot{\alpha}} = \begin{pmatrix}
		-p_3 -i p_4 & -p_1 +i p_2\\ -p_1 - i p_2 & p_3 - i p_4\\
	\end{pmatrix}
\ee
where we replace $\gamma_4$ by
\be
	\gamma_p = \tfrac{i}{2} v^{\dot{\alpha}}(p_{1 \dot{\alpha}}/z - p_{2 \dot{\alpha}})\,.
\ee
The corresponding spectral flow operation $\sigma_p$ is obtained by simply replacing $\gamma_4$ by $\gamma_p$.

It is straightforward to check that the $\sigma_p$ have the desired transformation rules with respect to scaling and rotations. For example, performing a scale transformation $x^\mu \to \Lambda x^\mu$ sends $v^{\dot{\alpha}} \to \Lambda v^{\dot{\alpha}}$ and therefore $J_a[f] \to J_a[\Lambda \cdot f]$, where
\be
	(\Lambda \cdot f)(v^{\dot{\alpha}}) = f(\Lambda^{-1} v^{\dot{\alpha}}) = \sum (\Lambda^{-(m_1+m_2)} f_{m_1,m_2}) (v^{\dot{1}})^{m_1} (v^{\dot{2}})^{m_2}
\ee
It follows that conjugating the spectral flow $\sigma_p$ by such a scale transformation $g_\Lambda$ results in $\sigma_{\Lambda^{-1} p}$
\be
	\sigma_{\Lambda^{-1} p} = g_\Lambda \sigma_p g_\Lambda{}^{-1}\,,
\ee
compatible with the expected scaling of momenta.

Rotations of the twistor sphere $z \to e^{i \theta} z$ are coupled with rotations of the fiber coordinates $v^{\dot{\alpha}} \to e^{i \theta/2} v_{\dot{\alpha}}$. Since these currents correspond to positive helicity gluons, they have spin $+1$ and so the generating function transforms as $J_a[\tilde{\lambda}](z_0) \to e^{-i \theta} J_a[e^{i \theta}\cdot f](z_0 e^{-i \theta})$, where
\be
	(e^{i \theta} \cdot f)(v^{\dot{\alpha}}) = f(e^{-i\theta/2} v^{\dot{\alpha}}) = \sum (e^{-i(m_1+m_2)\theta/2} f_{m_1,m_2}) (v^{\dot{1}})^{m_1} (v^{\dot{2}})^{m_2}
\ee
Thus, conjugating $\sigma_p$ by such a rotation is the same as rotating the momentum as $p_{1 \dot{\alpha}} \to e^{i \theta/2}p_{1 \dot{\alpha}}$ and $p_{2 \dot{\alpha}} \to e^{-i \theta/2}p_{2 \dot{\alpha}}$
\be
	\sigma_{e^{i \theta} p} = g_\theta \sigma_p g_\theta{}^{-1}\,.
\ee

\subsection{Pairwise helicity}
In the presence of the above monopole, the notion of helicity/spin of the gluons must be refined; this is reflected in the fact that $\sigma_p(J_a[m_1,m_2])$ does not have a definite 3d spin.

From a 4d perspective, an insertion of the current $J_a[m_1,m_2](z_0)$ corresponds to a positive helicity gluon with null momentum parameterized by the energy $\omega$ and the point $z_0$ on the twistor sphere $\PP^1$ via
\be
	p(\omega, z_0) = \omega \big(-(z_0 + \bar{z}_0), i(z_0 - \bar{z}_0), 1 - |z_0|^2, i(1+|z_0|^2)\big)\,.
\ee
If we consider the spectral flow operator for \emph{this} momentum (or any scalar multiple thereof), we find that $\gamma_p$ vanishes at the antipodal point $z = -1/z_0$ (and only at $z = -1/z_0$)
\be
	\gamma_{p(\omega,z_0)}(z) = i(z_0+z^{-1})\omega(\bar{z}_0 v^{\dot{2}} - v^{\dot{1}})\,.
\ee
In particular, the spectral flow $\sigma_{p(\omega,-1/z_0)}$ acts on the currents $J_a[f](z_0)$ in a particularly simple manner:
\be
\begin{aligned}
	\sigma_{p(\omega,-1/z_0)}(J_E[f])(z_0) & = z_0{}^{2} J_E[f](z_0)\\
	\sigma_{p(\omega,-1/z_0)}(J_H[f])(z_0) & = J_H[f](z_0)\\
	\sigma_{p(\omega,-1/z_0)}(J_F[f])(z_0) & = z_0{}^{-2} J_F[f](z_0)
\end{aligned}
\ee
We see that, in the presence of such a operator, the generating function $J_a[f](z_0)$ has an apparent shift in spin/helicity proportional to its weight with respect to the Cartan generator $J_H[0,0]$.

For the above, we see that the apparent shift in helicity is exactly reproduced by taking $q = \alpha \in \Lambda_{\rm root} \cong 2 \Z \subset \Lambda_{\rm weight} \cong \Z$ and $\m = 1 \in \Lambda_{\rm weight}^\vee \cong \Z$. In particular, we expect that correlation functions with this celestial monopole operator (obtained by the spectral flow $\sigma_p$) should correspond to scattering amplitudes in the presence of a 4d conformal primary of magnetic charge $\m = 1$, i.e. a 4d monopole. Moreover, scattering amplitudes involving states of non-trivial electro-magnetic charge (in the presence of some 4d local operator $\CO$) should be captured by correlation functions of the full, non-perturbative celestial chiral algebra (in the conformal block associated to $\CO$).

\subsection{Celestial OPE of gluons and monopoles}

Now that we have identified monopole operators in the celestial chiral algebra, e.g. states in the spectral flow of the vacuum module of the perturbative algebra, with magnetically charged scattering states, we expect that the OPEs of the currents $J, \tilde{J}$ and the operator $M_p$ associated to the state $\sigma_p(\ket{0})$ should encode the singularities of form factors involving scattering of gluons and a magnetic monopole with momentum $p$. We will be particularly interested in the case where the momentum $p = p(\omega, w)$ is null. Following the analysis of Section \ref{sec:monopolebundles}, it is straightforward to determine the 4d gauge field associated to such a null momentum:
\be
\begin{aligned}
	A_a & = \frac{1 + i (p \cdot x)(1+i \cot(p \cdot x))}{2 i p \cdot x}\big(\epsilon_{abc} \sigma^b p^c - p_4 \sigma_a\big)\\
	A_4 & = \frac{1 + i (p \cdot x)(1+i \cot(p \cdot x))}{2 i p \cdot x}\big(-p_a \sigma^a\big)
\end{aligned}
\ee
where, as above, $a = 1,2,3$. Note that this gauge field is singular at points $x \in \R^4$ where $p \cdot x$ is a non-zero integer multiple of $\pi$.

We will denote the corresponding spectral flow operation for the momentum $p(\omega, w)$ by $\sigma_w$. This spectral flow automorphism acts on the currents $J_a[m_1,m_2]$ as
\begin{equation}\label{eq:nullSF}
	\begin{aligned}
		\sigma_w(J_E[m_1,m_2](z_0)) & = \sum\limits_{l\geq 0} z_0{}^{2-l}\sum\limits_{m \geq 0} \sum\limits_{n = 0}^{l+m} \frac{(-2)^{l+m} \bar{w}^n w^m}{(l+m-n)! n!} \binom{l+m}{n} J_E[m_1+l+m-n,m_2+n](z_0)\\
		\sigma_w(J_H[m_1,m_2](z_0)) & = J_H[m_1,m_2](z_0)\\
		& \hspace{-1cm} - 2\sum\limits_{l\geq 0} z_0{}^{1-l}\sum\limits_{m \geq 0} \sum\limits_{n = 0}^{l+m} \frac{(-2)^{l+m} \bar{w}^n w^m}{(l+m+1)!} \binom{l+m}{n}\binom{l+m}{l} J_E[m_1+l+m-n,m_2+n](z_0)\\
		\sigma_w(J_F[m_1,m_2](z_0)) & = \sum\limits_{l\geq 0} z_0{}^{-2-l}\sum\limits_{m \geq 0} \sum\limits_{n = 0}^{l+m} \frac{2^{l+m} \bar{w}^n w^m}{(l+m-n)! n!} \binom{l+m}{n} J_F[m_1+l+m-n,m_2+n](z_0)\\
		& \hspace{-1cm} - 2\sum\limits_{l\geq 0} z_0{}^{-1-l}\sum\limits_{m \geq 0} \sum\limits_{n = 0}^{l+m} \frac{2^{l+m} \bar{w}^n w^m}{(l+m+1)!} \binom{l+m}{n}\binom{l+m}{l} J_H[m_1+l+m-n,m_2+n](z_0)\\
		& \hspace{-1cm} - 8\sum\limits_{l\geq 0} z_0{}^{-l}\sum\limits_{m \geq \lceil l/2 \rceil} \sum\limits_{n = 0}^{2m} \frac{2^{2m} \bar{w}^n w^{2m-l}}{(l+m+1)!} \binom{2m}{n}\binom{2m}{l} J_E[m_1+2m-n,m_2+n](z_0)
	\end{aligned}
\end{equation}
We will also need the inverse spectral flow automorphism $\sigma_w^{-1}$:
\begin{equation}\label{eq:invnullSF}
	\begin{aligned}
		\sigma_w^{-1}(J_E[m_1,m_2](z_0)) & = \sum\limits_{l\geq 0} z_0{}^{-2-l}\sum\limits_{m \geq 0} \sum\limits_{n = 0}^{l+m} \frac{2^{l+m} \bar{w}^n w^m}{(l+m-n)! n!} \binom{l+m}{n} J_E[m_1+l+m-n,m_2+n](z_0)\\
		\sigma_w^{-1}(J_H[m_1,m_2](z_0)) & = J_H[m_1,m_2](z_0)\\
		& \hspace{-1cm} - 2\sum\limits_{l\geq 0} z_0{}^{-1-l}\sum\limits_{m \geq 0} \sum\limits_{n = 0}^{l+m} \frac{2^{l+m} \bar{w}^n w^m}{(l+m+1)!} \binom{l+m}{n}\binom{l+m}{l} J_E[m_1+l+m-n,m_2+n](z_0)\\
		\sigma_w^{-1}(J_F[m_1,m_2](z_0)) & = \sum\limits_{l\geq 0} z_0{}^{2-l}\sum\limits_{m \geq 0} \sum\limits_{n = 0}^{l+m} \frac{(-2)^{l+m} \bar{w}^n w^m}{(l+m-n)! n!} \binom{l+m}{n} J_F[m_1+l+m-n,m_2+n](z_0)\\
		& \hspace{-1cm} - 2\sum\limits_{l\geq 0} z_0{}^{1-l}\sum\limits_{m \geq 0} \sum\limits_{n = 0}^{l+m} \frac{(-2)^{l+m} \bar{w}^n w^m}{(l+m+1)!} \binom{l+m}{n}\binom{l+m}{l} J_H[m_1+l+m-n,m_2+n](z_0)\\
		& \hspace{-1cm} - 8\sum\limits_{l\geq 0} z_0{}^{-l}\sum\limits_{m \geq \lceil l/2 \rceil} \sum\limits_{n = 0}^{2m} \frac{2^{2m} \bar{w}^n w^{2m-l}}{(l+m+1)!} \binom{2m}{n}\binom{2m}{l} J_E[m_1+2m-n,m_2+n](z_0)
	\end{aligned}
\end{equation}

We can now describe the OPE of the generators $J_a[\vec{m}]$ of the (perturbative part of the) celestial chiral algebra and the celestial monopole operator $V_w$ as before. The natural vacuum module is generated by the vacuum vector $\ket{0}$ that is annihilated by all non-negative modes of the currents $J_a[\vec{m}]$; the spectral flow module $\sigma_w(\CV)$ should then encode the celestial monopole operator $V_w$ (as well as those operators related to it by the action of the perturbative currents). 

Proceeding as in Section \ref{sec:bdymonopoles}, we find the following OPEs of the monopole operator $M_w(0)$ and the currents $J_a[\vec{m}]$. First, the OPE with $J_E[\vec{m}]$ is given by
\be
\begin{aligned}
	& J_E[m_1, m_2](z) V_w(0)\\
	& \sim \sum\limits_{l,m\geq 0} \sum\limits_{n = 0}^{l+m} \frac{2^{l+m} \bar{w}^n w^m}{(l+m-n)! n!} \binom{l+m}{n} \sum\limits_{k=0}^{l+1} z_0{}^{k-2-l} \bigg(\sigma_w\big(J_E[m_1+l+m-n, m_2+n]_{-k-1} \ket{0}\big)\bigg)(0).
\end{aligned}
\ee
Similarly, the OPE with the current $J_H[m_1, m_2]$ takes the form
\be
\begin{aligned}
	& J_H[m_1, m_2](z) V_w(0)\\
	& \sim - 2\sum\limits_{l,m \geq 0} \sum\limits_{n = 0}^{l+m} \frac{2^{l+m} \bar{w}^n w^m}{(l+m+1)!} \binom{l+m}{n}\binom{l+m}{l}\sum\limits_{k=0}^{l} z^{k-1-l}\bigg(\sigma_w\big(J_E[m_1+l+m-n,m_2+n]_{-k-1}\ket{0}\big)\bigg)(0)
\end{aligned}
\ee
and the OPE with the current $J_F[m_1, m_2]$ is
\be
\begin{aligned}
	& J_F[m_1, m_2](z) V_w(0)\\
	& \sim \sum\limits_{l\geq 3}\sum\limits_{m \geq 0} \sum\limits_{n = 0}^{l+m} \frac{(-2)^{l+m} \bar{w}^n w^m}{(l+m-n)! n!} \binom{l+m}{n} \sum\limits_{k=0}^{l-3} z_0{}^{k-l+2} \bigg(\sigma_w\big(J_F[m_1+l+m-n, m_2+n]_{-k-1} \ket{0}\big)\bigg)(0)\\
	& - 2\sum\limits_{l\geq 2}\sum\limits_{m \geq 0} \sum\limits_{n = 0}^{l+m} \frac{(-2)^{l+m} \bar{w}^n w^m}{(l+m+1)!} \binom{l+m}{n}\binom{l+m}{l} \sum\limits_{k=0}^{l-2}z_0{}^{k-l+1} \bigg(\sigma_w\big(J_H[m_1+l+m-n, m_2+n]_{-k-1} \ket{0}\big)\bigg)(0)\\
	& - 8\sum\limits_{l\geq 1}\sum\limits_{m \geq \lceil l/2 \rceil} \sum\limits_{n = 0}^{2m} \frac{2^{2m} \bar{w}^n w^{2m-l}}{(l+m+1)!} \binom{2m}{n}\binom{2m}{l} \sum\limits_{k=0}^{l-1}z_0{}^{k-l} \bigg(\sigma_w\big(J_E[m_1+2m-n, m_2+n]_{-k-1} \ket{0}\big)\bigg)(0)
\end{aligned}
\ee

The above discussion can be generalized to other gauge Lie algebras in a fairly straightforward manner. Namely, there is an analogous $\fg$ bundle on $\PT$ resulting in a magnetically charged excitation for any embedding $\fsl(2) \hookrightarrow \fg$ and hence spectral flow of the corresponding celestial current algebra. The gauge bundles arising from conjugate embeddings are equivalent to one another, as are the corresponding spectral flow modules.

It is an open problem, in both mathematics and physics, to understand how these families of spectral flow automorphisms, labeled by monopole momenta, help organize the module categories of perturbative celestial chiral algebras, as well as how these spectral flows play with the possible action of dualities.

\section{Conclusions \& Speculations}

We conclude by highlighting some major open questions and future directions raised by the investigation presented here. We have argued that it is fruitful to view celestial chiral algebras that admit twistorial uplifts as boundary chiral algebras of certain 3d holomorphic-topological theories. Our hope is that the 3d perspective, reviewed in Section \ref{sec:monopoles}, may shed some light on outstanding mysteries in the celestial holography program (at least for twistorial theories). 

In this work we have focused on studying certain (spectral flow) modules and OPEs of celestial chiral algebras, related to the addition of magnetically charged states in 3d and, variously, in 4d. The 4d and 3d perspectives on self-dual gauge theories are complementary, and isomorphic by the twistorial correspondence of \cite{CP22}, though as we have seen in this note it can be difficult to establish explicit isomorphisms between certain (magnetically charged) 3d and 4d states. In particular, in the case of self-dual QED for example, one important outstanding question is: what is the 4d interpretation of the spectral flows of the vacuum module $\bigoplus_m \sigma_m |0 \rangle$? As described in Section \ref{sec:4dinterpret}, these modules are charged under the symmetry generated by the Goldstone field $S[0,0]$ canonically conjugate to the generator $J[0,0]$ of the conformally soft photon theorem, which is distinct from Strominger's soft magnetic photon theorem. In a sense, these modules are canonically conjugate to the Faddeev-Kulish dressing factors, cf. \cite{Arkani-Hamed:2020gyp, Choi:2018oel}, which are gauge-theoretic analogs of the modules described recently in \cite{Crawley:2023brz}, Better understanding these operators will be the key step towards applying our 3d algebraic reasoning to potential celestial holographic interpretations or, more generally, the implications for asymptotic symmetries in flat spacetime. 

There are a natural line defects in the self-dual abelian gauge theory, cf. Section \ref{sec:selfdualmonopole}. We expect to them realize yet more modules for the celestial chiral algebra, but these are still out of reach. The main technical obstruction arises when one lacks a description of the corresponding 6d bundles on twistor space that is amenable to the requisite sheaf (or Dolbeault) cohomology computation. Perhaps these bundles can be studied explicitly in relative cohomology, following Bailey \cite{bailey1985twistors} or using some form of tangential Cauchy Riemann cohomology as in \cite{Zeng2023}; we leave such computations to the future.

Another important element of this correspondence is the map between local operators in 4d and conformal blocks of the celestial chiral algebra. A choice of conformal block is necessary in order to recast simple tree-level scattering amplitudes of electric and magnetically charged states as correlators in our chiral algebras, which depend on the choice of conformal block. The space of conformal blocks (on a surface $\Sigma$) of a vertex algebra of boundary local operators is expected to be realized as Hilbert space of the bulk 3d theory (on $\Sigma$). These state spaces are relatively well understood in simple examples of twisted 3d $\CN=2$ theories, cf. \cite{Bullimore:2018yyb}, but they have yet to be described in the examples of interest to this paper. In future work, we plan to study the spaces of conformal blocks relevant to our abelian examples in the presence of nontrivial monopole bundles, realized as holomorphic bundles on twistor space, and the corresponding spaces of states, particularly in the context of self-dual QED where we can use the analysis in \cite{Zeng:2021zef} and where a free-field realization of the celestial chiral algebra is possible.

Another fascinating application of the 3d perspective is the possibility to generate conjectures about nonperturbative enhancements of/corrections to the (classical) asymptotic symmetry algebras of conformally soft modes by magnetically charged states. In the abelian case, we showed that the extended celestial chiral algebra can be realized as a simple current extension, being a direct sum over spectral flow modules. In the non-abelian case, following the result for ordinary Chern-Simons theory at positive level, it is natural to anticipate from 3d reasoning that the nonperturbative celestial chiral algebra can be realized as a simple current extension of a quotient module, where one removes singular vectors. However, the analogous result for zero-level Chern-Simons theory, which arises in the celestial setting, has not yet been determined; moreover, in the case of zero-level abelian Chern-Simons theory, using the standard vertex algebra inner product, \textit{all} states in the perturbative chiral algebra are null although the vacuum module character is non-vanishing. From a spacetime point of view, there are singular vectors in the perturbative algebra, and their removal is crucial for the implementation of Ward identities for soft theorems \cite{Banerjee:2019aoy, Banerjee:2019tam}, but it is important to note that the relevant conjugation operation on the celestial currents is \emph{not} the same as the natural one for a vertex algebra and instead reflects the pairing of incoming and outgoing scattering states, cf. Section 3 of \cite{Pasterski:2021fjn}. The definition of this pairing for integer conformal dimension, which is the case for the twistorial celestial chiral algerbas discussed in this paper, is particularly subtle and still being developed \cite{Freidel:2022skz}.

We expect that many of these issues can be resolved by a more concrete description of 3d non-abelian monopoles supplementing the abstract proposal of \cite{CDG}. This is an active area of research in derived algebraic geometry and in the representation theory of vertex algebras. It is our hope that such mathematical conjectures will ultimately be translated, via the twistorial correspondence, into concrete selection rules for four-dimensional physics -- with a better understanding of the boundary chiral algebras for non-abelian gauge theories as well as their conformal blocks, the perspective of \cite{CP22} offers a direct window into the scattering of 4d states with general electric and magnetic charges.

For twistorial theories, one can also characterize celestial chiral algebras at the quantum level \cite{CPassoc, Fernandez:2023abp,  Bittleston:2022jeq}. It would be interesting to explore quantum deformations of the full nonperturbative chiral algebras (i.e. including 3d boundary monopole operators); such a characterization will necessarily include the generators $E[r,s], F[r,s]$, associated to the anomaly-cancelling axion of \cite{Costello:2021bah, CP22} which restores associativity. At the classical level, these generators are gauge-invariant and should be unchanged by the spectral flow morphism.

The twistorial studies discussed above are in fact recent contributions in a long history of work exploring the quantization of holomorphic BF and Chern-Simons theories on twistor (and ambitwistor) space, recently summarized in \cite{Geyer:2022cey}; see also \cite{Geyer:2014lca} for earlier connections to asymptotic symmetries. It will be fascinating to flesh out further connections between twisted holography, celestial holography, and (ambi)twistor string theory, as well as the 3d holomorphic-topological perspective espoused in this note.

Finally, it will be very interesting to explore, from this 3d perspective, possible nonperturbative enhancements of the celestial chiral algbebra for self-dual gravity, whose twistorial uplift has been recently studied in \cite{Bittleston:2022jeq, Bittleston:2022nfr}; for further progress on understanding the celestial symmetries of gravity from a twistorial point of view, see \cite{Adamo:2021lrv, Mason:2022hly}. For instance, the twistorial ``quadrille'' construction for self-dual dyons reviewed in \ref{sec:selfdualmonopole} has long been known to admit Schwarzschild or Kerr-like analogues in the nonlinear graviton theory of Penrose \cite{sparling1976non} (they can be thought of perhaps more precisely as self-dual Taub-NUT solutions, with a fixed relationship between the Schwarzschild mass and NUT charge). It would be very interesting to study the 3d HT and chiral algebraic interpretations of such configurations along the lines of this work, and perhaps connect to the recent work of \cite{Crawley:2023brz, Guevara:2021yud}; the higher-spin multipole moments encoded in the charges under the $w_{1 + \infty}$ algebra for the self-dual black hole solutions studied in \textit{loc. cit.} may admit a geometric realization in terms of relative cohomology in a non-Hausdorff twistor space. 

\acknowledgments
We are grateful to K. Costello, T. Creutzig, S. Pasterski, and especially L. Mason and A. Sharma for very helpful conversations and correspondences. NP also thanks Perimeter Institute for hospitality while this work was being completed. NG and NP are supported by funds from the Department of Physics and the College of Arts \& Sciences at the  University of Washington. NP is also supported by the DOE Early Career
Research Program under award DE-SC0022924, and by the Visiting Fellowship program at the Perimeter Institute of Theoretical Physics. Research at Perimeter Institute is supported by the Government of Canada through Industry Canada and by the
Province of Ontario through the Ministry of Research and Innovation.  

\appendix

\section{Twistor Space Conventions}
\label{app:twistor}
Twistor space $\PT$ is the total space of the vector bundle $\CO(1) \oplus \CO(1) \to \PP^1$, where $\CO(1) \to \PP^1$ is the line bundle on $\PP^1$ whose sections transform homogeneously with weight 1; i.e. if $z_\alpha$ are homogeneous coordinates on $\PP^1$ and $\sigma(z_\alpha)$ is a section of $\CO(1) \to \PP^1$, then $\sigma(\lambda z_\alpha) = \lambda \sigma(z_\alpha)$. It can be viewed as an open subset of $\PP^3$ as follows. Let $Z^A$ be homogeneous coordinates on $\PP^3$, and split them as $Z^A = (\mu^{\dot{\alpha}}, z_\alpha)$, where $\alpha = 1,2$ and $\dot{\alpha} = \dot{1}, \dot{2}$. Twistor space $\PT$ can then be identified with the open set $z_\alpha \neq 0$; the $z_\alpha$ are identified with homogeneous coordinates on $\PP^1$ and $\mu^{\dot{\alpha}}$ fiber coordinates.

We will often work with the affine coordinate $z = z_1/z_2$ on $\PP^1$ and corresponding fiber coordinates $v^{\dot{\alpha}} = \mu^{\dot{\alpha}}/z_2$. Under the transition function to the other affine patch $z \to w = z_2/z_1 = 1/z$, we find that the fiber coordinates transform as $v^{\dot{\alpha}} \to u^{\dot{\alpha}} = \mu_{\dot{\alpha}}/z_1 = v^{\dot{\alpha}}/z$. To summarize:
\be
\begin{aligned}
	w & = 1/z \qquad & u^{\dot{\alpha}} & = v^{\dot{\alpha}}/z\,,\\
	z & = 1/w \qquad & v^{\dot{\alpha}} & = u^{\dot{\alpha}}/w\,.
\end{aligned}
\ee

For every point $x^\mu \in \C^4$ there is a (linearly) embedded $\PP^1$ given by the following \textit{incidence relation}:
\be
	\mu^{\dot{\alpha}} = x^\mu (\sigma_\mu)^{\dot{\alpha} \alpha}z_\alpha\,,
\ee
where the $\sigma_a$ are essentially the Pauli matrices
\be
	(\sigma_a)^{\dot{\alpha} \alpha} = \Bigg\{ \begin{pmatrix}
		0 & 1\\ 1 & 0\\
	\end{pmatrix}\,,\quad \begin{pmatrix}
		0 & -i\\ i & 0\\
	\end{pmatrix}\,,\quad \begin{pmatrix}
		1 & 0\\ 0 & -1\\
	\end{pmatrix}\,,\quad \begin{pmatrix}
		-i & 0\\ 0 & -i\\
	\end{pmatrix}
	\Bigg\}\,.
\ee
With this choice, we find that the determinant of the matrix
\be
	x^{\dot{\alpha} \alpha} = x^\mu (\sigma_\mu)^{\dot{\alpha} \alpha} = \begin{pmatrix}
		x^3 - i x^4 & x^1 - i x^2\\ x^1 + i x^2 & -x^3 - i x^4\\
	\end{pmatrix}
\ee
is the negative of the Euclidean norm, $\det x = -\sum_\mu (x^\mu)^2$. In terms of the affine coordinates, the above incidence relations takes the following form:
\be
\begin{aligned}
	v^{\dot{1}} & = x^2 + i x^1 + z(x^4 + i x^3)\,, \qquad & v^{\dot{2}} & = x^4 - i x^3 + z(-x^2 + i x^1)\\
	u^{\dot{1}} & = w(x^4 + i x^3) + x^2 + i x^1\,, \qquad & u^{\dot{2}} & = w(-x^2 + i x^1) + x^4 - i x^3\\
\end{aligned}\,.
\ee
We follow the convention that $SU(2)$ indices are raised and lowered as $X^\beta = X_\alpha \epsilon^{\alpha \beta}$ and $X_\alpha = \epsilon_{\alpha \beta}X^\beta$ with $\epsilon_{12} = 1 = \epsilon^{12}$. A standard argument shows that restricting to real $x^\mu$, this gives us an identification
\be
	\PT \cong \PP^1 \times \R^4
\ee
as \emph{real} manifolds.

\subsection{Rotating twistor space}

The action of the Euclidean rotation group $SO(4) \cong (SU(2)_+ \times SU(2)_-)/\Z_2$ is represented as usual: the index $\alpha$ (resp. $\dot{\alpha}$) is an index for the fundamental representation of $SU(2)_-$ (resp. $SU(2)_+$).

The diagonal torus in $SU(2)_-$ sends the homogeneous coordinate $z_\alpha = [z:1]$ on $\PP^1$ to $[e^{-i\theta/2} z, e^{i\theta/2}]$ and hence acts as a rotation of the $z$ plane $z \to z e^{-i \theta}$ together with a rotation of the fiber coordinates $v^{\dot{\alpha}} \to e^{-i \theta/2} v^{\dot{\alpha}}$. In terms of our real foliation, we see that this acts on the $\alpha$ index of the coordinates $x^{\dot{\alpha} \alpha}$ as expected -- the $x^{\dot{\alpha}1}$ rotate as $x^{\dot{\alpha} 1} \to e^{i \theta/2}x^{\dot{\alpha} 1}$ and the $x^{\dot{\alpha}2}$ rotate as $x^{\dot{\alpha} 2} \to e^{-i \theta/2}x^{\dot{\alpha} 2}$.


\bibliographystyle{JHEP}
\bibliography{celestial}

\providecommand{\href}[2]{#2}\begingroup\raggedright\begin{thebibliography}{10}

\bibitem{Guevara:2021abz}
A.~Guevara, E.~Himwich, M.~Pate and A.~Strominger, \emph{{Holographic symmetry
  algebras for gauge theory and gravity}},
  \href{https://doi.org/10.1007/JHEP11(2021)152}{\emph{JHEP} {\bfseries 11}
  (2021) 152} [\href{https://arxiv.org/abs/2103.03961}{{\ttfamily
  2103.03961}}].

\bibitem{Strominger:2021lvk}
A.~Strominger, \emph{{w(1+infinity) and the Celestial Sphere}},
  \href{https://arxiv.org/abs/2105.14346}{{\ttfamily 2105.14346}}.

\bibitem{Ball:2021tmb}
A.~Ball, S.~A. Narayanan, J.~Salzer and A.~Strominger, \emph{{Perturbatively
  exact w$_{1+\infty}$ asymptotic symmetry of quantum self-dual gravity}},
  \href{https://doi.org/10.1007/JHEP01(2022)114}{\emph{JHEP} {\bfseries 01}
  (2022) 114} [\href{https://arxiv.org/abs/2111.10392}{{\ttfamily
  2111.10392}}].

\bibitem{Bhardwaj:2022anh}
R.~Bhardwaj, L.~Lippstreu, L.~Ren, M.~Spradlin, A.~Yelleshpur~Srikant and
  A.~Volovich, \emph{{Loop-level gluon OPEs in celestial holography}},
  \href{https://arxiv.org/abs/2208.14416}{{\ttfamily 2208.14416}}.

\bibitem{Monteiro:2022lwm}
R.~Monteiro, \emph{{Celestial chiral algebras, colour-kinematics duality and
  integrability}},  \href{https://arxiv.org/abs/2208.11179}{{\ttfamily
  2208.11179}}.

\bibitem{CP22}
K.~Costello and N.~M. Paquette, \emph{{Celestial holography meets twisted
  holography: 4d amplitudes from chiral correlators}},
  \href{https://arxiv.org/abs/2201.02595}{{\ttfamily 2201.02595}}.

\bibitem{CPassoc}
K.~Costello and N.~M. Paquette, \emph{{On the associativity of one-loop
  corrections to the celestial OPE}},
  \href{https://arxiv.org/abs/2204.05301}{{\ttfamily 2204.05301}}.

\bibitem{Fernandez:2023abp}
V.~E. Fern\'andez, \emph{{One-loop corrections to the celestial chiral algebra
  from Koszul Duality}},  \href{https://arxiv.org/abs/2302.14292}{{\ttfamily
  2302.14292}}.

\bibitem{Bittleston:2022jeq}
R.~Bittleston, \emph{{On the associativity of 1-loop corrections to the
  celestial operator product in gravity}},
  \href{https://doi.org/10.1007/JHEP01(2023)018}{\emph{JHEP} {\bfseries 01}
  (2023) 018} [\href{https://arxiv.org/abs/2211.06417}{{\ttfamily
  2211.06417}}].

\bibitem{Costello:2021bah}
K.~J. Costello, \emph{{Quantizing local holomorphic field theories on twistor
  space}},  \href{https://arxiv.org/abs/2111.08879}{{\ttfamily 2111.08879}}.

\bibitem{Costello:2019jsy}
K.~Costello and S.~Li, \emph{{Anomaly cancellation in the topological string}},
  \href{https://doi.org/10.4310/ATMP.2020.v24.n7.a2}{\emph{Adv. Theor. Math.
  Phys.} {\bfseries 24} (2020) 1723}
  [\href{https://arxiv.org/abs/1905.09269}{{\ttfamily 1905.09269}}].

\bibitem{Costello:2022jpg}
K.~Costello, N.~M. Paquette and A.~Sharma, \emph{{Top-down holography in an
  asymptotically flat spacetime}},
  \href{https://arxiv.org/abs/2208.14233}{{\ttfamily 2208.14233}}.

\bibitem{Costello:2018zrm}
K.~Costello and D.~Gaiotto, \emph{{Twisted Holography}},
  \href{https://arxiv.org/abs/1812.09257}{{\ttfamily 1812.09257}}.

\bibitem{CPKoszul}
K.~Costello and N.~M. Paquette, \emph{{Twisted Supergravity and Koszul Duality:
  A case study in AdS$_3$}},
  \href{https://doi.org/10.1007/s00220-021-04065-3}{\emph{Commun. Math. Phys.}
  {\bfseries 384} (2021) 279}
  [\href{https://arxiv.org/abs/2001.02177}{{\ttfamily 2001.02177}}].

\bibitem{Bakalov_2003}
B.~Bakalov and V.~Kac, \emph{{Field algebras}},
  \href{https://doi.org/10.1155/s1073792803204232}{\emph{International
  Mathematics Research Notices} {\bfseries 2003} (2003) 123}.

\bibitem{Strominger:2015bla}
A.~Strominger, \emph{{Magnetic Corrections to the Soft Photon Theorem}},
  \href{https://doi.org/10.1103/PhysRevLett.116.031602}{\emph{Phys. Rev. Lett.}
  {\bfseries 116} (2016) 031602}
  [\href{https://arxiv.org/abs/1509.00543}{{\ttfamily 1509.00543}}].

\bibitem{Kapec:2021eug}
D.~Kapec and P.~Mitra, \emph{{Shadows and soft exchange in celestial CFT}},
  \href{https://doi.org/10.1103/PhysRevD.105.026009}{\emph{Phys. Rev. D}
  {\bfseries 105} (2022) 026009}
  [\href{https://arxiv.org/abs/2109.00073}{{\ttfamily 2109.00073}}].

\bibitem{Bershadsky:1993cx}
M.~Bershadsky, S.~Cecotti, H.~Ooguri and C.~Vafa, \emph{{Kodaira-Spencer theory
  of gravity and exact results for quantum string amplitudes}},
  \href{https://doi.org/10.1007/BF02099774}{\emph{Commun. Math. Phys.}
  {\bfseries 165} (1994) 311}
  [\href{https://arxiv.org/abs/hep-th/9309140}{{\ttfamily hep-th/9309140}}].

\bibitem{ACMV}
M.~Aganagic, K.~Costello, J.~McNamara and C.~Vafa, \emph{Topological
  chern-simons/matter theories},
  \href{https://arxiv.org/abs/1706.09977}{{\ttfamily 1706.09977}}.

\bibitem{Aharony:1997bx}
O.~Aharony, A.~Hanany, K.~A. Intriligator, N.~Seiberg and M.~J. Strassler,
  \emph{{Aspects of N=2 supersymmetric gauge theories in three-dimensions}},
  \href{https://doi.org/10.1016/S0550-3213(97)00323-4}{\emph{Nucl. Phys. B}
  {\bfseries 499} (1997) 67}
  [\href{https://arxiv.org/abs/hep-th/9703110}{{\ttfamily hep-th/9703110}}].

\bibitem{Intriligator:2013lca}
K.~Intriligator and N.~Seiberg, \emph{{Aspects of 3d N=2 Chern-Simons-Matter
  Theories}}, \href{https://doi.org/10.1007/JHEP07(2013)079}{\emph{JHEP}
  {\bfseries 07} (2013) 079} [\href{https://arxiv.org/abs/1305.1633}{{\ttfamily
  1305.1633}}].

\bibitem{Gadde:2013wq}
A.~Gadde, S.~Gukov and P.~Putrov, \emph{{Walls, Lines, and Spectral Dualities
  in 3d Gauge Theories}},
  \href{https://doi.org/10.1007/JHEP05(2014)047}{\emph{JHEP} {\bfseries 05}
  (2014) 047} [\href{https://arxiv.org/abs/1302.0015}{{\ttfamily 1302.0015}}].

\bibitem{Gadde:2013sca}
A.~Gadde, S.~Gukov and P.~Putrov, \emph{{Fivebranes and 4-manifolds}},
  \href{https://doi.org/10.1007/978-3-319-43648-7_7}{\emph{Prog. Math.}
  {\bfseries 319} (2016) 155}
  [\href{https://arxiv.org/abs/1306.4320}{{\ttfamily 1306.4320}}].

\bibitem{Okazaki:2013kaa}
T.~Okazaki and S.~Yamaguchi, \emph{{Supersymmetric boundary conditions in
  three-dimensional N=2 theories}},
  \href{https://doi.org/10.1103/PhysRevD.87.125005}{\emph{Phys. Rev. D}
  {\bfseries 87} (2013) 125005}
  [\href{https://arxiv.org/abs/1302.6593}{{\ttfamily 1302.6593}}].

\bibitem{DGP}
T.~Dimofte, D.~Gaiotto and N.~M. Paquette, \emph{{Dual boundary conditions in
  3d SCFT\textquoteright{}s}},
  \href{https://doi.org/10.1007/JHEP05(2018)060}{\emph{JHEP} {\bfseries 05}
  (2018) 060} [\href{https://arxiv.org/abs/1712.07654}{{\ttfamily
  1712.07654}}].

\bibitem{CDG}
K.~Costello, T.~Dimofte and D.~Gaiotto, \emph{{Boundary Chiral Algebras and
  Holomorphic Twists}},  \href{https://arxiv.org/abs/2005.00083}{{\ttfamily
  2005.00083}}.

\bibitem{Zeng:2021zef}
K.~Zeng, \emph{{Monopole Operators and Bulk-Boundary Relation in Holomorphic
  Topological Theories}},  \href{https://arxiv.org/abs/2111.00955}{{\ttfamily
  2111.00955}}.

\bibitem{BDGHK16}
M.~Bullimore, T.~Dimofte, D.~Gaiotto, J.~Hilburn and H.-C. Kim, \emph{{Vortices
  and Vermas}}, \href{https://doi.org/10.4310/ATMP.2018.v22.n4.a1}{\emph{Adv.
  Theor. Math. Phys.} {\bfseries 22} (2018) 803}
  [\href{https://arxiv.org/abs/1609.04406}{{\ttfamily 1609.04406}}].

\bibitem{Nak}
H.~Nakajima, \emph{{Towards a mathematical definition of Coulomb branches of
  $3$-dimensional $\mathcal{N}=4$ gauge theories, I}},
  \href{https://doi.org/10.4310/ATMP.2016.v20.n3.a4}{\emph{Adv. Theor. Math.
  Phys.} {\bfseries 20} (2016) 595}
  [\href{https://arxiv.org/abs/1503.03676}{{\ttfamily 1503.03676}}].

\bibitem{BFN}
A.~Braverman, M.~Finkelberg and H.~Nakajima, \emph{{Towards a mathematical
  definition of Coulomb branches of $3$-dimensional $\mathcal{N} = 4$ gauge
  theories, II}}, \href{https://doi.org/10.4310/ATMP.2018.v22.n5.a1}{\emph{Adv.
  Theor. Math. Phys.} {\bfseries 22} (2018) 1071}
  [\href{https://arxiv.org/abs/1601.03586}{{\ttfamily 1601.03586}}].

\bibitem{BDG}
M.~Bullimore, T.~Dimofte and D.~Gaiotto, \emph{{The Coulomb Branch of 3d
  ${\mathcal{N}= 4}$ Theories}},
  \href{https://doi.org/10.1007/s00220-017-2903-0}{\emph{Commun. Math. Phys.}
  {\bfseries 354} (2017) 671}
  [\href{https://arxiv.org/abs/1503.04817}{{\ttfamily 1503.04817}}].

\bibitem{ADLslab}
S.~Alekseev, M.~Dedushenko and M.~Litvinov, \emph{{Chiral life on a slab}},
  \href{https://arxiv.org/abs/2301.00038}{{\ttfamily 2301.00038}}.

\bibitem{Zhu}
X.~Zhu, \emph{{An introduction to the affine Grassmannian and the geometric
  Satake equivalence}},  \href{https://arxiv.org/abs/1603.05593}{{\ttfamily
  1603.05593}}.

\bibitem{BCDN23}
A.~Ballin, T.~Creutzig, T.~Dimofte and W.~Niu, \emph{{3d mirror symmetry of
  braided tensor categories}},
  \href{https://arxiv.org/abs/2304.11001}{{\ttfamily 2304.11001}}.

\bibitem{BN22}
A.~Ballin and W.~Niu, \emph{{3d Mirror Symmetry and the $\beta\gamma$ VOA}},
  \href{https://arxiv.org/abs/2202.01223}{{\ttfamily 2202.01223}}.

\bibitem{GN23}
N.~Garner and W.~Niu, \emph{{Line Operators in $U(1|1)$ Chern-Simons Theory}},
  \href{https://arxiv.org/abs/2304.05414}{{\ttfamily 2304.05414}}.

\bibitem{CR13}
T.~Creutzig and D.~Ridout, \emph{{W-Algebras Extending Affine
  $\mathfrak{gl}(1|1)$}},
  \href{https://doi.org/10.1007/978-4-431-54270-4_24}{\emph{Springer Proc.
  Math. Stat.} {\bfseries 36} (2013) 349}
  [\href{https://arxiv.org/abs/1111.5049}{{\ttfamily 1111.5049}}].

\bibitem{CKM17}
T.~Creutzig, S.~Kanade and R.~McRae, \emph{{Tensor categories for vertex
  operator superalgebra extensions}},
  \href{https://arxiv.org/abs/1705.05017}{{\ttfamily 1705.05017}}.

\bibitem{CMY22}
T.~Creutzig, R.~McRae and J.~Yang, \emph{{Direct limit completions of vertex
  tensor categories}},
  \href{https://doi.org/10.1142/S0219199721500334}{\emph{Commun. Contemp.
  Math.} {\bfseries 24} (2022) 2150033}
  [\href{https://arxiv.org/abs/2006.09711}{{\ttfamily 2006.09711}}].

\bibitem{GW2018}
O.~Gwilliam and B.~R. Williams, \emph{{Higher Kac\textendash{}Moody algebras
  and symmetries of holomorphic field theories}},
  \href{https://doi.org/10.4310/ATMP.2021.v25.n1.a4}{\emph{Adv. Theor. Math.
  Phys.} {\bfseries 25} (2021) 129}
  [\href{https://arxiv.org/abs/1810.06534}{{\ttfamily 1810.06534}}].

\bibitem{Zeng2023}
K.~Zeng, \emph{{Twisted Holography and Celestial Holography from Boundary
  Chiral Algebra}},  \href{https://arxiv.org/abs/2302.06693}{{\ttfamily
  2302.06693}}.

\bibitem{Donnay:2022sdg}
L.~Donnay, S.~Pasterski and A.~Puhm, \emph{{Goldilocks modes and the three
  scattering bases}},
  \href{https://doi.org/10.1007/JHEP06(2022)124}{\emph{JHEP} {\bfseries 06}
  (2022) 124} [\href{https://arxiv.org/abs/2202.11127}{{\ttfamily
  2202.11127}}].

\bibitem{Freidel:2022skz}
L.~Freidel, D.~Pranzetti and A.-M. Raclariu, \emph{{A discrete basis for
  celestial holography}},  \href{https://arxiv.org/abs/2212.12469}{{\ttfamily
  2212.12469}}.

\bibitem{Mason:2010yk}
L.~J. Mason and D.~Skinner, \emph{{The Complete Planar S-matrix of N=4 SYM as a
  Wilson Loop in Twistor Space}},
  \href{https://doi.org/10.1007/JHEP12(2010)018}{\emph{JHEP} {\bfseries 12}
  (2010) 018} [\href{https://arxiv.org/abs/1009.2225}{{\ttfamily 1009.2225}}].

\bibitem{Adamo:2011pv}
T.~Adamo, M.~Bullimore, L.~Mason and D.~Skinner, \emph{{Scattering Amplitudes
  and Wilson Loops in Twistor Space}},
  \href{https://doi.org/10.1088/1751-8113/44/45/454008}{\emph{J. Phys. A}
  {\bfseries 44} (2011) 454008}
  [\href{https://arxiv.org/abs/1104.2890}{{\ttfamily 1104.2890}}].

\bibitem{Bu:2022dis}
W.~Bu and E.~Casali, \emph{{The 4d/2d correspondence in twistor space and
  holomorphic Wilson lines}},
  \href{https://doi.org/10.1007/JHEP11(2022)076}{\emph{JHEP} {\bfseries 11}
  (2022) 076} [\href{https://arxiv.org/abs/2208.06334}{{\ttfamily
  2208.06334}}].

\bibitem{OhZhou}
J.~Oh and Y.~Zhou, \emph{{Twisted holography of defect fusions}},
  \href{https://doi.org/10.21468/SciPostPhys.10.5.105}{\emph{SciPost Phys.}
  {\bfseries 10} (2021) 105}
  [\href{https://arxiv.org/abs/2103.00963}{{\ttfamily 2103.00963}}].

\bibitem{GaiottoOh}
D.~Gaiotto and J.~Oh, \emph{{Aspects of $\Omega$-deformed M-theory}},
  \href{https://arxiv.org/abs/1907.06495}{{\ttfamily 1907.06495}}.

\bibitem{GaiottoRapcak}
D.~Gaiotto and M.~Rapcak, \emph{{Miura operators, degenerate fields and the
  M2-M5 intersection}},
  \href{https://doi.org/10.1007/JHEP01(2022)086}{\emph{JHEP} {\bfseries 01}
  (2022) 086} [\href{https://arxiv.org/abs/2012.04118}{{\ttfamily
  2012.04118}}].

\bibitem{Zwanziger:1972sx}
D.~Zwanziger, \emph{{Angular distributions and a selection rule in charge-pole
  reactions}}, \href{https://doi.org/10.1103/PhysRevD.6.458}{\emph{Phys. Rev.
  D} {\bfseries 6} (1972) 458}.

\bibitem{Csaki:2020inw}
C.~Csaki, S.~Hong, Y.~Shirman, O.~Telem, J.~Terning and M.~Waterbury,
  \emph{{Scattering amplitudes for monopoles: pairwise little group and
  pairwise helicity}},
  \href{https://doi.org/10.1007/JHEP08(2021)029}{\emph{JHEP} {\bfseries 08}
  (2021) 029} [\href{https://arxiv.org/abs/2009.14213}{{\ttfamily
  2009.14213}}].

\bibitem{Csaki:2022tvb}
C.~Cs\'aki, Z.-Y. Dong, O.~Telem, J.~Terning and S.~Yankielowicz,
  \emph{{Dressed vs. pairwise states, and the geometric phase of monopoles and
  charges}}, \href{https://doi.org/10.1007/JHEP02(2023)211}{\emph{JHEP}
  {\bfseries 02} (2023) 211}
  [\href{https://arxiv.org/abs/2209.03369}{{\ttfamily 2209.03369}}].

\bibitem{CHSTTW20}
C.~Csaki, S.~Hong, Y.~Shirman, O.~Telem, J.~Terning and M.~Waterbury,
  \emph{{Scattering amplitudes for monopoles: pairwise little group and
  pairwise helicity}},
  \href{https://doi.org/10.1007/JHEP08(2021)029}{\emph{JHEP} {\bfseries 08}
  (2021) 029} [\href{https://arxiv.org/abs/2009.14213}{{\ttfamily
  2009.14213}}].

\bibitem{vanBeest:2023dbu}
M.~van Beest, P.~Boyle~Smith, D.~Delmastro, Z.~Komargodski and D.~Tong,
  \emph{{Monopoles, Scattering, and Generalized Symmetries}},
  \href{https://arxiv.org/abs/2306.07318}{{\ttfamily 2306.07318}}.

\bibitem{Donnay:2018neh}
L.~Donnay, A.~Puhm and A.~Strominger, \emph{{Conformally Soft Photons and
  Gravitons}}, \href{https://doi.org/10.1007/JHEP01(2019)184}{\emph{JHEP}
  {\bfseries 01} (2019) 184}
  [\href{https://arxiv.org/abs/1810.05219}{{\ttfamily 1810.05219}}].

\bibitem{Crawley:2023brz}
E.~Crawley, A.~Guevara, E.~Himwich and A.~Strominger, \emph{{Self-Dual Black
  Holes in Celestial Holography}},
  \href{https://arxiv.org/abs/2302.06661}{{\ttfamily 2302.06661}}.

\bibitem{Kulish:1970ut}
P.~P. Kulish and L.~D. Faddeev, \emph{{Asymptotic conditions and infrared
  divergences in quantum electrodynamics}},
  \href{https://doi.org/10.1007/BF01066485}{\emph{Theor. Math. Phys.}
  {\bfseries 4} (1970) 745}.

\bibitem{Chung:1965zza}
V.~Chung, \emph{{Infrared Divergence in Quantum Electrodynamics}},
  \href{https://doi.org/10.1103/PhysRev.140.B1110}{\emph{Phys. Rev.} {\bfseries
  140} (1965) B1110}.

\bibitem{Arkani-Hamed:2020gyp}
N.~Arkani-Hamed, M.~Pate, A.-M. Raclariu and A.~Strominger, \emph{{Celestial
  amplitudes from UV to IR}},
  \href{https://doi.org/10.1007/JHEP08(2021)062}{\emph{JHEP} {\bfseries 08}
  (2021) 062} [\href{https://arxiv.org/abs/2012.04208}{{\ttfamily
  2012.04208}}].

\bibitem{Choi:2018oel}
S.~Choi and R.~Akhoury, \emph{{Soft Photon Hair on Schwarzschild Horizon from a
  Wilson Line Perspective}},
  \href{https://doi.org/10.1007/JHEP12(2018)074}{\emph{JHEP} {\bfseries 12}
  (2018) 074} [\href{https://arxiv.org/abs/1809.03467}{{\ttfamily
  1809.03467}}].

\bibitem{sparling1977dynamically}
G.~Sparling, \emph{Dynamically broken symmetry and global yang-mills in
  minkowski space}, {\emph{Further Advances in Twistor Theory} {\bfseries 1}
  (1977) 171}.

\bibitem{PS79}
R.~Penrose and G.~Sparling, \emph{{The Twistor Quadrille: A Line Bundle Based
  on the Coulomb Field}},  in \emph{Advances in Twistor Theory}, L.~J. Mason,
  L.~P. Hughston, P.~Z. Kobak and K.~Pulverer, eds., CRC Press, (1979).

\bibitem{WWtwistor}
R.~S. Ward and R.~O. Wells, \emph{{Twistor geometry and field theory}},
  Cambridge Monographs on Mathematical Physics. Cambridge University Press, 8,
  1991,
  \href{https://doi.org/10.1017/CBO9780511524493}{10.1017/CBO9780511524493}.

\bibitem{Guevara:2021yud}
A.~Guevara, \emph{{Reconstructing Classical Spacetimes from the S-Matrix in
  Twistor Space}},  \href{https://arxiv.org/abs/2112.05111}{{\ttfamily
  2112.05111}}.

\bibitem{bailey1985twistors}
T.~Bailey, \emph{Twistors and fields with sources on worldlines},
  {\emph{Proceedings of the Royal Society of London. A. Mathematical and
  Physical Sciences} {\bfseries 397} (1985) 143}.

\bibitem{Bullimore:2018yyb}
M.~Bullimore and A.~Ferrari, \emph{{Twisted Hilbert Spaces of 3d Supersymmetric
  Gauge Theories}}, \href{https://doi.org/10.1007/JHEP08(2018)018}{\emph{JHEP}
  {\bfseries 08} (2018) 018}
  [\href{https://arxiv.org/abs/1802.10120}{{\ttfamily 1802.10120}}].

\bibitem{Banerjee:2019aoy}
S.~Banerjee, P.~Pandey and P.~Paul, \emph{{Conformal properties of soft
  operators: Use of null states}},
  \href{https://doi.org/10.1103/PhysRevD.101.106014}{\emph{Phys. Rev. D}
  {\bfseries 101} (2020) 106014}
  [\href{https://arxiv.org/abs/1902.02309}{{\ttfamily 1902.02309}}].

\bibitem{Banerjee:2019tam}
S.~Banerjee and P.~Pandey, \emph{{Conformal properties of soft-operators. Part
  II. Use of null-states}},
  \href{https://doi.org/10.1007/JHEP02(2020)067}{\emph{JHEP} {\bfseries 02}
  (2020) 067} [\href{https://arxiv.org/abs/1906.01650}{{\ttfamily
  1906.01650}}].

\bibitem{Pasterski:2021fjn}
S.~Pasterski, A.~Puhm and E.~Trevisani, \emph{{Celestial diamonds: conformal
  multiplets in celestial CFT}},
  \href{https://doi.org/10.1007/JHEP11(2021)072}{\emph{JHEP} {\bfseries 11}
  (2021) 072} [\href{https://arxiv.org/abs/2105.03516}{{\ttfamily
  2105.03516}}].

\bibitem{Geyer:2022cey}
Y.~Geyer and L.~Mason, \emph{{The SAGEX review on scattering amplitudes Chapter
  6: Ambitwistor Strings and Amplitudes from the Worldsheet}},
  \href{https://doi.org/10.1088/1751-8121/ac8190}{\emph{J. Phys. A} {\bfseries
  55} (2022) 443007} [\href{https://arxiv.org/abs/2203.13017}{{\ttfamily
  2203.13017}}].

\bibitem{Geyer:2014lca}
Y.~Geyer, A.~E. Lipstein and L.~Mason, \emph{{Ambitwistor strings at null
  infinity and (subleading) soft limits}},
  \href{https://doi.org/10.1088/0264-9381/32/5/055003}{\emph{Class. Quant.
  Grav.} {\bfseries 32} (2015) 055003}
  [\href{https://arxiv.org/abs/1406.1462}{{\ttfamily 1406.1462}}].

\bibitem{Bittleston:2022nfr}
R.~Bittleston, A.~Sharma and D.~Skinner, \emph{{Quantizing the non-linear
  graviton}},  \href{https://arxiv.org/abs/2208.12701}{{\ttfamily 2208.12701}}.

\bibitem{Adamo:2021lrv}
T.~Adamo, L.~Mason and A.~Sharma, \emph{{Celestial $w_{1+\infty}$ Symmetries
  from Twistor Space}},
  \href{https://doi.org/10.3842/SIGMA.2022.016}{\emph{SIGMA} {\bfseries 18}
  (2022) 016} [\href{https://arxiv.org/abs/2110.06066}{{\ttfamily
  2110.06066}}].

\bibitem{Mason:2022hly}
L.~Mason, \emph{{Gravity from holomorphic discs and celestial $Lw_{1+\infty}$
  symmetries}},  \href{https://arxiv.org/abs/2212.10895}{{\ttfamily
  2212.10895}}.

\bibitem{sparling1976non}
G.~Sparling, \emph{The non-linear graviton representing the analogue of
  schwarzschild or kerr black holes}, {\emph{Twistor Newslett} {\bfseries 1}
  (1976) 14}.

\end{thebibliography}\endgroup
	
\end{document}